\documentclass{aa}  

\usepackage{epstopdf}
\usepackage{graphicx}
\usepackage{txfonts}
\usepackage{lineno}

\begin{document}

   \title{Reinvestigating $\alpha$~Cen~AB in light of asteroseismic forward and inverse methods}


   \author{S.J.A.J. Salmon
          \inst{1,2}
          \and
          V. Van Grootel \inst{1}
          \and
          G. Buldgen \inst{2}
          \and
          M-A. Dupret \inst{1}
          \and
          P. Eggenberger \inst{2}
          }

   \institute{STAR Institute, Universit\'e de Li\`ege, All\'ee du 6 Ao\^ut 19C, 4000 Li\`ege, Belgium \\
              \email{sebastien.salmon@uliege.be}
         \and
              Observatoire de Gen\`eve, Universit\'e de Gen\`eve, 51 Ch. des Maillettes, 1290, Sauverny, Switzerland }

   \date{Received September 15, 1996; accepted March 16, 1997}

 
  \abstract
   {The $\alpha$~Cen stellar system is the closest neighbour to our Sun. Its main component is a binary composed of two main-sequence stars, one  more  massive than the Sun and one less massive. The system's  bright magnitude  led to a wealth of astronomical observations over a long period, making it an appealing testbed for stellar physics. In particular, detection of stellar pulsations in both $\alpha$~Cen~A and B has revealed the potential of asteroseismology for determining its fundamental stellar parameters. Asteroseismic studies have also focused on the presence of a convective core in the A component, but as yet without definitive confirmation.}
   {Progress in the determination of solar surface abundances and stellar opacities have yielded new input for stellar theoretical models. We investigate their impact on a reference system such as $\alpha$~Cen~AB. We seek to confirm the presence of a convective core in $\alpha$~Cen~A by analysing the role of different stellar physics and the potential of asteroseismic inverse methods.}
   {First, we present a new series of asteroseismic calibrations of the binary carried out using forward approach modelling and 
including updated chemical mixture and opacities in the models. We took advantage of the most up-to-date orbital 
solution as non-seismic constraints. We then complement our analysis with help of recent asteroseismic diagnostic tools 
based on inverse methods developed for solar-like stars.}
   {The inclusion of an updated chemical mixture -that is less metal-rich- appears to reduce the predicted asteroseismic masses 
of each component. Neither classical asteroseismic indicators such as frequency ratios, nor asteroseismic inversions 
favour the presence of a convective core in $\alpha$~Cen~A. The quality of the observational seismic dataset is the main 
limiting factor to settle the issue. Implementing new observing strategies to improve the precision on the pulsation 
frequencies would certainly refine the outcome of asteroseismology for this binary system. } 
   {}

   \keywords{Asteroseismology -- Stars: solar-type -- binaries: general -- Stars: oscillations
               }
               
 \titlerunning{Reinvestigating $\alpha$~Cen~AB}

\maketitle
%

\section{Introduction}

The knowledge and characterisation of solar-type stars is a dominant subject of 
modern stellar physics. They are the best candidates to shelter 
planetary systems favourable to the development of life, while they are also key for a comparative study of 
the evolution and structure of our Sun. Solar-type stars can   be precisely characterised  as they exhibit pressure modes of pulsations, stochastically excited by near-surface convection. Interpreting the  pulsations with asteroseismology provides tight constraints on the fundamental parameters of these stars and reveals their internal structure.
 
The CoRoT \citep{baglin06} and Kepler \citep{borucki10} space missions confirmed this potential, thanks to the data of 
unprecedented quality they delivered. This naturally 
benefited the study of solar-type stars \citep[e.g.][]{metcalfe12,chaplin13,lebreton14,sa15,buldgen16a,buldgen16b,lund17}.
Before this golden age for asteroseismology, exploitable detections of solar-like pulsations from the ground were only
possible for a few stars \citep{bedding08}. Among them, $\alpha$ Cen was and remains of particular interest for the 
study of solar-type stars. It is composed of two dwarf G- and K-type stars forming a binary system, of eccentricity e$\sim$0.52 and orbital period P$\sim$80 yr \citep{pourbaix16}, and a red dwarf, Proxima, thought to be gravitationally bound 
\citep{kervella17b}. The masses of the primary and secondary components are respectively estimated to 
M$_{\textrm{A}}\sim$1.11 M$_{\odot}$ and M$_{\textrm{B}}\sim$0.94 M$_{\odot}$ \citep{kervella16}. Consequently, the two 
stars are very similar to the Sun and are privileged targets for understanding the physics of solar-type stars. In particular, 
$\alpha$~Cen~A is at the very limit in mass for the onset of convection in its nuclear-burning core. It thus represents 
an excellent testbed to the formalism of heat transport by convection, which remains a flaw of current stellar 
models. 

As the closest stellar system to the Sun, the $\alpha$ Cen system presents bright apparent magnitudes of 
$V_{\textrm{A}}$=0.01 and $V_{\textrm{B}}$=1.33 \citep{ducati02}, making it easier to perform a thorough analysis by 
numerous observing campaigns. Most were done for stellar physics purposes, resulting in a rich set of tight constraints 
on its stellar components, from exploiting the follow-up of their orbital motion 
\citep[e.g.][]{wesselink53,kamper78,pourbaix02,pourbaix16,kervella16} to determining their stellar atmospheres 
\citep[e.g.][]{edvardsson88,neuforge97,bigot08,portodemello08,morel18}. Renewed interest in the system came with the 
bloom of exoplanetology research.  While the stability and existence of a habitable zone for exoplanets in the 
$\alpha$~Cen~AB system has been theorised \citep[e.g.][]{kaltenegger13,quarles16}, claims for the detection of a planet 
orbiting $\alpha$~Cen~B \citep{dumusque12} have been dismissed \citep{hatzes13,rajpaul16} or await further evidence 
\citep{demory15}. Pursuing the characterisation of this system hence remains a crucial stake for stellar physics and 
the search for exoplanets. 

A privileged approach is combining non-seismic and seismic information to constrain to the best level possible the fundamental parameters 
and the history of $\alpha$~Cen~AB. Pulsations were expected to be present in the system components because of 
their similarity with the Sun. Before their observational detection was possible, the properties of putative 
$\alpha$~Cen stellar pulsations had thus already been studied \citep[e.g.][]{edmonds92,kim99,neuforge99,guenther00}. 
Alpha~Cen~A was   the first solar-type star for which oscillations were detected from ground-based 
observations \citep{kjeldsen99,bouchy01,bouchy02,bedding04} and from space by the WIRE spacecraft 
\citep{schou01,fletcher06}. After the first asteroseismic modellings of $\alpha$ Cen~A \citep{thevenin02,thoul03}, 
  oscillations in $\alpha$~Cen~B \citep{carrier03,kjeldsen05} were also reported, opening the way to 
asteroseismic modelling of the binary system 
\citep[e.g.][]{eggenberger04,miglio05,yildiz07,yildiz08,tang08,joyce18}. The first series of asteroseismic 
studies \citep{eggenberger04,miglio05,yildiz07} showed a good agreement with masses obtained from the orbital solution 
and parallax estimate \citep[e.g.][]{pourbaix02} and interferometric radii \citep{kervella03}. Since then the picture 
of the system and stellar physics has  evolved. Using longer observational runs, \citet{pourbaix16} revised their 
estimate of the masses of $\alpha$~Cen~A and B by respectively about 3\% and 5\%. Between the first asteroseismic 
studies, important revision of the solar chemical composition \citep{asplund05,asplund09} and stellar opacities 
\citep{colgan16} were also proposed. Moreover, several asteroseismic studies investigated whether $\alpha$~Cen~A 
harbours a convective core \citep{miglio05,demeulenaer10,bazot12,bazot16,nsamba18,nsamba19}. None could firmly confirm 
its presence, and instead result in contradictory tendencies, depending on the asteroseismic dataset, namely suggesting 
no convective core in \citet{demeulenaer10}, favouring one in \citet{nsamba19}.

In this work we investigate the consequences on the asteroseismic modelling of recent developments in the physics of 
stellar models and check their consistency with the revised orbital solution of $\alpha$~Cen~AB. We look at the impact 
of the metallicity scale of reference, which is generally the solar chemical mixture.  
The abundances of the elements that describe the solar mixture are derived mostly from the observation of the Sun's 
photosphere. The measure of abundances from the photospheric spectra is not a side issue. For instance, 
\citet{asplund05} and \citet{asplund09} derived, using  a new approach for the analysis of solar spectra, a severe decrease in the solar 
metallicity ($\sim$30\%) in comparison to previous works \citep[e.g.][]{grevesse93}. This modification of the 
metallicity reference itself will have an obvious impact on the internal properties of stellar models, in particular the opacity of the 
stellar plasma \citep[e.g.][]{buldgen19}. The revision of the solar metallicity also led to a 
disagreement between theoretical models of the Sun and helioseismology \citep[e.g.][]{bahcall05}. This solar issue raises 
the question of the accuracy of stellar physics ingredients of models, with particular concern for the opacity. In response, 
thorough numerical computations of opacity data for solar and stellar astrophysics were re-initiated 
\citep[e.g.][]{mondet15,lepennec15,colgan16}. In this framework we implement the recent Los Alamos opacities \citep{colgan16} 
and assess whether asteroseismic calibrations of $\alpha$~Cen~AB are sensible to the opacity changes.

The observational dataset of $\alpha$~Cen also benefited from new investigations. We take advantage of the 
asteroseismic dataset derived by \citet{demeulenaer10}, which by combining multi-site observing campaigns has revealed 
the highest number and most precise oscillation frequencies to date of   $\alpha$~Cen~A.  We hence calibrated a series 
of new models based on this frequency set and using \citet{asplund09} as the metallicity scale. We compared our results 
with the new orbital constraints from \citet{pourbaix16} and \citet{kervella16}, and interferometric radii from 
\citet{kervella17}. The models, which are calibrated in a forward modelling approach, are then used as references to 
infer the presence of a convective core in $\alpha$~Cen~A. We performed this with the  help of asteroseismic inversions, based on the 
innovative framework developed in \citet{buldgenmeand,buldgenS}. Our inversions are constructed to retrieve the entropy 
in the central layers of the stars, whose value behaves in a clear-cut way between the regime of convection or pure 
radiative transport.

We start by presenting in section 2 the non-seismic and seismic constraints, and the input stellar physics we used for the 
asteroseismic calibrations of $\alpha$~Cen~AB. Section~3 describes the calibration method by forward approach and recalls 
the basis for seismic inversion of the entropy. In Sect.~4 we present the results of our forward asteroseismic 
calibrations and discuss the impact on them of the revised metallicity scale of reference. In Sect.~5 we perform the seismic 
inversions and discuss their potential for assessing the presence of a convective core. We present our concludions in Sect.~6.


\section{Framework for asteroseismic calibrations of $\alpha$~Cen~AB}
\label{observables}

\subsection{Non-seismic observables}
\label{observables-ns}

The $\alpha$~Cen~AB orbital parameters were recently revised in two studies by \citet[][hereafter P16]{pourbaix16} and 
\citet[][hereafter K16]{kervella16}, which are today the works of reference on this topic. As the system is a 
double-lined spectroscopic and visual binary, we can constrain from astrometry and spectroscopic radial velocities the total mass (M$_{\textrm{A}}$+M$_{\textrm{B}}$) and the fractional mass 
M$_{\textrm{B}}$/(M$_{\textrm{A}}$+M$_{\textrm{B}}$), hereafter denoted by $\kappa$, of the system. With knowledge of 
the parallax, it is then possible to disentangle the individual masses. However, the parallax is a parameter adjusted to the 
same level as the fractional mass when solving the orbital system of equations \citep[e.g.][]{pourbaix98}. The 
individual masses in that approach are hence implicitly dependent on the other orbital parameters, both in P16 and K16. 
As noted by K16, the  parallax determination of each team explains most of the differences between the 
individual masses determined by these two studies. We see in Table \ref{paramorbital} that their values for $\kappa$ 
differ by 0.6\%, but rise to $\sim$3\% for the individual masses. To avoid any dependence on the parallax derivation 
and its associated uncertainties, we adopt the fractional mass rather than the individual masses as observational 
constraints in most of our calibrations. The values of P16 and K16 for $\kappa$ are very similar, and we only selected 
the  K16 value because their analysis included a longer follow-up of the  astrometric record. 

The radius ratio R$_{\textrm{A}}$/R$_{\textrm{B}}$ (denoted $\Gamma$) is obtained from angular diameters 
determined by interferometry and is an interesting parameter to exploit. It is indeed firmly constrained because  of 
its independence of the wavelength band of the observation and of the parallax. From a new set of interferometric data, 
\citet{kervella17} derived a value $\Gamma=$1.4172 $\pm$ 0.0016, in perfect agreement with their earlier estimation 
\citep{kervella03}. We consequently chose it as the second observational constraints for our calibrations. Nevertheless, we 
also   performed some with the individual radii and masses from \citet{kervella17} and K16 (the same parallax is adopted 
in both studies) to analyse how they influence the results of our calibrations (see Sect. \ref{section4}).

\begin{table*}
\caption{Non-seismic observational parameters for $\alpha$~Cen~AB.}             
\label{paramorbital}      
\centering          
\begin{tabular}{l l l}     
\hline\hline       
Fundamental parameter &  &  Reference \\
\hline
\hline
Fractional mass ($\kappa$) & M$_{\textrm{B}}$/(M$_{\textrm{A}}$+M$_{\textrm{B}}$)$=0.4617 \pm 0.0004$ & P16 \\
&  M$_{\textrm{B}}$/(M$_{\textrm{A}}$+M$_{\textrm{B}}$)$=0.45884 \pm 0.00027$ & K16 \\
\hline
Individual masses  & M$_{\textrm{A}}= 1.133 \pm 0.005 \ \ \ \textrm{M}_{\textrm{B}} = 0.972 \pm 0.005 $ & P16 \\
  & M$_{\textrm{A}}= 1.1055 \pm 0.0039 \ \ \ \textrm{M}_{\textrm{B}} = 0.9373 \pm 0.0033 $ & K16 \\
\hline
Radius ratio ($\Gamma$) & $\textrm{R}_{\textrm{A}}/\textrm{R}_{\textrm{B}}=1.4172 \pm 0.0016$ & \citet{kervella17} \\
\hline
Individual radii & $\textrm{R}_{\textrm{A}}=1.2234 \pm 0.0053 \ \ \ \textrm{R}_{\textrm{B}}=0.8632 \pm 0.0037$ & 
\citet{kervella17} \\
\hline
Effective temperature & $\textrm{T}_{\textrm{eff,A}}=5795 \pm 19 \textrm{K} \ \ \ \textrm{T}_{\textrm{eff,B}}=5231 \pm 
21 \textrm{K}$ & \citet{kervella17} \\
\hline
Luminosity & $\textrm{L}_{\textrm{A}}=1.521 \pm 0.015$ \ \ \ $\textrm{L}_{\textrm{B}}=0.503 \pm 0.007$ & 
\citet{kervella17}\\
\hline
 Metallicity& [Fe/H]$_{\textrm{A}}=0.25 \pm 0.02$ \ \ \ [Fe/H]$_{\textrm{B}}=0.24 \pm 0.03$ & \citet{neuforge97} \\
& [Fe/H]$_{\textrm{A}}=0.24 \pm 0.03$ \ \ \ [Fe/H]$_{\textrm{B}}=0.25 \pm 0.04$ & \citet{portodemello08} \\
& [Fe/H]$_{\textrm{A}}=0.16 \pm 0.05$ \ \ \ [Fe/H]$_{\textrm{B}}=-$ & \citet{bigot08} \\
& [Fe/H]$_{\textrm{A}}=0.24 \pm 0.01$ \ \ \ [Fe/H]$_{\textrm{B}}=0.22\pm0.02$ & \citet{morel18} \\
\hline
 \end{tabular}
\tablefoot{Masses, radii, and luminosities are expressed in corresponding solar units.}

\end{table*}

The system is well known for being metal rich \citep{french71} and, owing to its brightness, it has motivated many 
spectroscopic studies of its stellar component atmospheres. Discrepancies exist between studies \citep[see e.g. 
discussion in][]{portodemello08}, but most of the metallicity determinations of $\alpha$~Cen~AB give a value for the 
system of [Fe/H]$\sim$0.25 \citep[see the short review in][]{morel18}. Used in the first series of asteroseismic 
studies, the reference values of \citet{neuforge97} are in excellent agreement with more the recent determinations of 
\citet{portodemello08} and \citet{morel18} (see Table \ref{paramorbital}). As in \citet{neuforge97}, the work of 
\citet{portodemello08} found a negligible difference between the A and B component metallicities. Therefore, we selected 
from these latter authors the single [Fe/H]=0.24 value as the common present-day surface metallicity to be reproduced 
by our models for each component. Despite convergence between the different spectroscopic studies (based on 1D stellar 
atmosphere models), \citet{bigot08} found a departing value, [Fe/H]=0.16, that they derived for $\alpha$~Cen~A with 3D 
hydrodynamical simulations.   Although the authors mention this work as preliminary, their value is intriguing and 
deserves to be considered. We also carry out some calibrations of the system using this lower metallicity value.

Effective temperatures (T$_{\textrm{eff}}$) derived in the spectroscopic studies are in the ranges 
$5750 \lesssim$T$_{\textrm{eff,A}} \lesssim 5850$~K and $5150 \lesssim$T$_{\textrm{eff,B}} \lesssim 5300$~K  for the A 
and B components \citep[see review in][]{morel18}. Based on the bolometric fluxes of the stars, 
\citet[][and references therein]{kervella17} derived effective temperatures (see Table \ref{paramorbital}) in perfect 
agreement with the spectroscopic values. We choose their values as constraints because we also use the fractional mass and 
radius ratio derived by the same authors.

\begin{table}[!]
\caption{Sets of observational constraints to calibrate models}             
\label{table-explicite-models}      
\centering          
\begin{tabular}{l l l }     
\hline\hline       
Model name & Observational constraints &  
Reference
\\
includes:  & & 
\\
\hline
\hline
$\Delta_\nu \delta_\nu$ & $\Delta \nu_{\textrm{A}}$  \ \ \ $\Delta \nu_{\textrm{B}}$ (lin. 
regr. on 
$\nu_{\ell=0,1,2,3}$) & dM10  \\
 & $\langle \delta \nu_{02,\textrm{A}}\rangle$ \ \ \ $\langle\delta \nu_{02,\textrm{B}}\rangle $ & 
 \\
\hline
r$_{10}$ & individual r$_{10}$(n) of each star & dM10  \\
\hline
r$_{02}$ & individual r$_{02}$(n) of each star & dM10  \\
\hline
r$_{13}$ & individual r$_{13}$(n) of each star & dM10  \\
\hline
R,T$_{\textrm{eff}}$& R$_{\textrm{A}}$, R$_{\textrm{B}}$ & 
K17 \\   
& T$_{\textrm{eff},A}^\dagger$, T$_{\textrm{eff},B}^\dagger$ & K17 \\
& Z/X$|_\textrm{S}=0.0307 \pm 0.0021^{\ddag}$ & PdM08 \\
\hline
 M$_{\rm frac}$,R$_{\rm ratio}$ & M$_{\textrm{B}}$/(M$_{\textrm{A}}$+M$_{\textrm{B}}$) & K16\\
 & R$_{\textrm{A}}$/R$_{\textrm{B}}$ & K17 \\
 & L$_{\textrm{A}}$,L$_{\textrm{B}}$ & K17 \\ 
 & Z/X$|_\textrm{S}=0.0307 \pm 0.0021^{\ddag}$ & PdM08 \\
 \hline
 M,R & M$_{\textrm{A}}$, M$_{\textrm{B}}$ & K16$^+$\\
 & R$_{\textrm{A}}$,R$_{\textrm{B}}$ & 
K17 \\
 & L$_{\textrm{A}}$,L$_{\textrm{B}}$ & K17 \\ 
 & Z/X$|_\textrm{S}=0.0307 \pm 0.0021^{\ddag}$ & PdM08 \\
 \hline
  Z/X$_{\textrm{BTK08}}$ & Z/X$|_\textrm{S}=0.0255 \pm 0.0018^*$ & BTK08 \\ 
 \hline
 -GN93 & Z/X$|_\textrm{S}=0.0415 \pm 0.0029^{\maltese}$ & PdM08 \\
 \hline 
\end{tabular}
\tablefoot{If not given here, values of the constraints can be found in Table \ref{paramorbital}.\\ Key to references: 
dM10 \citep{demeulenaer10}; K16 \citep{kervella16} ; K17 \citep{kervella17}; $^\dagger$ 
K17 derived 
$\sigma_{\textrm{T}_{\textrm{eff}}} \sim20$K, we instead adopted a conservative error of 60 K; $^{\ddag}$ assuming 
the AGSS09 solar distribution of elements and [Fe/H]$=+0.24 \pm 0.03$ from PdM08 \citep{portodemello08}; $^+$ K16 derived 
$\sigma_{\textrm{M}} \sim$ 0.003 M$_{\odot}$; we revised the error to 0.03 M$_{\odot}$ to enable the unbiased 
performance of our local optimisation method (see Sect.~\ref{asteroseismicmethods}); $^*$ assuming the  AGSS09 solar 
distribution of elements and [Fe/H]$=+0.16 \pm 0.05$ from BTK08 \citep{bigot08}; $^{\maltese}$ assuming the GN93 
solar distribution of elements and [Fe/H]$=+0.24 \pm 0.03$ from PdM08.}

\end{table}

\begin{table}[!]
\caption{Input physics of the calibrations}             
\label{table-explicite-models-2}      
\centering          
\begin{tabular}{l l l l l}     
\hline\hline       
Name & Chemical & Opacity & $\alpha_{\textrm{ov}}$ & Diffusion
\\
includes:  & mixture & & &  \\
\\
\hline
\hline
 & AGSS09 & OPAL &  & Yes  \\
 \hline
-GN93  & GN93 & OPAL &  & Yes \\
\hline
-OPLIB & AGSS09 & OPLIB &  & Yes \\
\hline 
-Ov & & & 0.10 & \\
\hline 
-Ov0.20 & & & 0.20 & \\
\hline
-Surf & \multicolumn{4}{l}{Oscillation surface effects} \\
& \multicolumn{4}{l}{are corrected following} \\
 & \multicolumn{4}{l}{\citet{sonoi15}} \\
\hline
\end{tabular}
\tablefoot{Key to references: K16 \citep{kervella16}; K17 \citep{kervella17}; $^\dagger$ K16 derived 
$\sigma_{\textrm{T}_{\textrm{eff}}} \sim20$K, we instead adopt a conservative error of 60 K; $^{\ddag}$ assuming 
the AGSS09 solar distribution of elements and [Fe/H]$=+0.24 \pm 0.03$ from PdM08 \citep{portodemello08}.}
\end{table}

\subsection{Seismic constraints}
\label{sectionseismic}
Oscillations in $\alpha$~Cen~A were analysed by \citet{bouchy01} and \citet{bouchy02} from a 13-night radial velocity 
dataset, resulting in the detection of 28 modes (with angular degrees $\ell=0,1,2$). Independently, \citet{bedding04} 
and \citet{butler04} observed the star for five nights, detecting 42 modes ($\ell=0,1,2,3$). More recently, \citet{bazot07} 
detected 34 modes ($\ell=0-3$) from a new five-night run of observations. Notably, the dataset from \citet{bouchy02}, 
and \citet{bedding04} and \citet{butler04}, were taken on overlapping dates in 2001, from multi-site facilities (in 
Chile and Australia). \citet{demeulenaer10}, in an effort to combine those data, reduced the aliases induced by the 
day/night duty cycle. They obtained the most precise and complete set of acoustic oscillations for $\alpha$ Cen A, with 
46 modes from $\ell=0$ to 3. For the frequency set of $\alpha$~Cen~B, we adopt the richest one derived by 
\citet{kjeldsen05}, consisting of 37 acoustic modes with $\ell=0$ to 3.

The oscillation frequencies can be combined to define seismic indicators sensitive to different stellar properties. We considered in particular the large and small differences of mode frequencies, $\Delta~\nu_{n,\ell}$ and 
$\delta~\nu_{n,\ell}$, defined as
\begin{equation}
\label{eq1}
 \Delta \nu_{n,\ell}=\nu_{n,\ell}-\nu_{n-1,\ell}
\end{equation}
and
\begin{equation}
\label{eq2}
 \delta \nu_{n,\ell}=\nu_{n,\ell}-\nu_{\textrm{n-1},\ell+2},
\end{equation}
where $\nu_{n,\ell}$ is the frequency of a mode of radial order $n$ and angular degree $\ell$. It is a good approximation to assume the acoustic 
oscillations of solar-like stars are in the asymptotic regime. In this regime it can be shown 
\citep[e.g.][]{Gough2003} that these frequency differences approaches constant values known as the large and small 
separations, which are proportional to structural stellar quantities. The latter is a proxy of the mean stellar 
density, the former depends on the gradient of sound speed, which is mostly sensitive to the chemical stratification in the 
stellar core.

When comparison to observations are made,  particular care is required to retrieve these indicators from the 
theoretical models in a similar way. For instance, the observed large separations for $\alpha$ Cen are derived from the 
autocorrelation of the asymptotic formula to the oscillation spectra  \citep[e.g.][]{bouchy02}, which is hardly 
reproducible without bias from theoretical frequencies. Instead, we derived the observational large separation $\Delta 
\nu$ from a linear fit to the individual frequencies ($\nu$ as a function of $n$) for each $\ell$ and $n$ 
detected. We then computed the weighted mean of the four fitted values ($\ell=0,1,2,3$), obtaining $\Delta\nu_{\textrm{A}}=105.9~\pm~0.3\ 
\mu$Hz and $\Delta\nu_{\textrm{B}}=161.4~\pm~0.3\ \mu$Hz, respectively for the A and B component. Uncertainties were 
taken as the standard errors of each estimate and then propagated. We implemented in our calibrations the calculation of 
the theoretical large separations similarly, i.e. from linear fits to the theoretical frequencies of the models, on the same $\ell$ and $n$ as observed.
For the small separation we worked with the arithmetic mean value of the small differences for $\ell=0$, 
$\delta\nu_{n,0}$, following the definition in Eq.~\ref{eq2}. The observational values are then 
$\langle\delta\nu_{0,\textrm{A}}\rangle=5.63~\pm~0.73~\mu$Hz and 
$\langle\delta\nu_{0,\textrm{B}}\rangle=10.90~\pm~1.85~\mu$Hz. The theoretical values were computed similarly. 

These two indicators are combined to constrain the mean density of solar-like oscillators and their evolutionary 
stage, but they can suffer a bias due to surface effects affecting the observed individual frequencies. A solution is to 
add surface effect corrective terms to theoretical frequencies \citep[e.g.][]{kjeldsen08}. The form of the corrections and 
how they are  computed  are still a matter of intense debate. Another approach consists in defining seismic indicators 
as insensitive to surface effects as possible. \citet{roxburgh03} proposed to divide the small frequency separations by 
the large ones to break free from these effects:
\begin{equation}
\label{eq3}
r_{10}(n)=\frac{d_{10}(n)}{\Delta \nu_{n+1,0}} ; r_{02}(n)=\frac{\delta \nu_{n,0}}{\Delta \nu_{n,1}} ; 
r_{13}=\frac{\delta \nu_{n,1}}{\Delta \nu_{n+1,0}}.
\end{equation}
Here $d_{10}$ is the five-point small separation, as defined in Eq.~5 of \citet{roxburgh03}. These indicators are 
sensitive to variations in the chemical composition profile in the central stellar layers. They provide insightful information on 
the age and the physical conditions in the nuclear core. Their potential was tested in detail for $\alpha$~Cen~A in 
\citet{miglio05} and \citet{demeulenaer10} as posterior constraints. The $r_{10}$ and $r_{13}$ ratios seemed promising 
indicators of the energy transport process in the central layers, by showing a distinct behaviour between models of 
$\alpha$~Cen~A with or without convective core. We pursued this effort by including indicators of Eq.~\ref{eq3} as 
{a priori} seismic constraints for a series of calibrations.

\subsection{Physics of the models}

Relying on the observational constraints described above, we computed all our numerical 
models with the Li\`{e}ge stellar evolution code \citep[CLES,][]{cles}. The treatment of convection follows the 
mixing-length prescription \citep{bohm58}, and is implemented following \citet{cox68}. Except for some 
calibrations (see Table~\ref{table-explicite-models}), no overshooting was 
considered. When included, overshooting is implemented as an instantaneous extra-mixing extending over a region whose 
size is $\alpha_{\textrm{ov}} \times \min[r_{cc},H_P(r_{cc})]$, with $r_{cc}$ the size of the convective core, $H_P$ 
the local pressure scale height, and $\alpha_{\textrm{ov}}$ an overshooting parameter. The surface boundary conditions 
are obtained from Eddington's law ($T[\tau]$, $T$ being the temperature) 
for a grey atmosphere, with the atmosphere extended down to an optical depth $\tau \sim 10^{-6}$. The nuclear reaction 
rates were those of the Nacre (for nuclei with atomic mass A>15) and Nacre-II (A<16) compilations \citep{angulo99,xu13}, 
with the exception of the two first reactions of the pp-I chain whose rates were taken from 
\citet{adelberger11}. We adopted the updated solar chemical mixture of \cite{asplund09} (hereafter AGSS09). Opacities 
corresponding to this chemical mixture are computed with the OPAL tables \citep{iglesias96}, completed at low temperatures by opacities from 
\citet{ferguson05}. Similarly, the equation of state is computed with the FreeEOS code \citep{irwin12}. All our models included microscopic diffusion with resolution of Burgers' equations following the routine of \citet{thoul94}. The diffusion procedure considers three elements, H, He, and Fe, every 
element heavier than He being assimilated to Fe.

The choice of the AGSS09 mixture and its implications on the asteroseismic inferences we obtain must be further 
commented. The new and improved determination of the solar photospheric abundances made by \citet{asplund09}
led to a significant decrease in the solar metallicity, which was independently confirmed (albeit to a lower extent) by 
\cite{caffau11}. This severe decrease in  metallicity opened an important issue of modern stellar physics as the 
once near perfect agreement between helioseismology and solar theoretical models broke down \citep[e.g.][]{basu08}. To 
reconcile this agreement, the most obvious solutions would require an increase in the metallicity in the envelope of the 
Sun or an increase in the opacity. Although we selected the most up-to-date stellar physics for our models, 
this  solar stalemate still raises questions regarding  the interpretation of the solutions. Studies of $\alpha$~Cen prior to 2005 obviously used an old 
solar chemical mixture. The same is true for more recent studies performed by \citet{bazot12,bazot16}, which used models from \citet[][hereafter GN93]{grevesse93}, while \citet{nsamba18} used those of \citet{grevesse98}. However, \citet{miglio05} and \citet{demeulenaer10} used models computed either with GN93 or \citet{asplund05} and illustrated the importance of this choice for the age and emergence of a convective core in the models they inferred 
from asteroseismology. Consequently, we also tested the importance of the chemical mixture by adopting the GN93 mixture 
in some of our calibrations, as indicated in Table~\ref{table-explicite-models}.

The validity of the opacity data used in stellar evolution models, as mentioned above, is often invoked as a good 
candidate to solve the solar issue. It has resulted in new more advanced computations of opacity for stellar physics 
purpose by the Los Alamos group \citep[OPLIB]{colgan16}. These OPLIB opacities were developed independently to the 
widely used OPAL opacity tables. We thus  carried out some of our calibrations with OPLIB to assess 
whether they change the inferred stellar parameters. 

Finally, independently of those choices, all the adiabatic frequencies of oscillations of our models were obtained with 
the stellar pulsation code LOSC \citep{losc}. We did not apply any treatment for surface effects, except for one 
calibration for which we made the correction to the frequencies proposed by \citet{sonoi15} (see Sect. 
\ref{section4}).

\section{Asteroseismic methods}
\label{asteroseismicmethods}
Based on the latest observational constraints on $\alpha$~Cen~AB described in  Sect.~\ref{observables}, we first 
investigated the consequences on the inferences we can obtain with asteroseismology. We started by 
deriving a set of asteroseismic models in a forward approach similar to that in \citet{miglio05}. We then used the 
models forwardly obtained as the reference models for asteroseismic structural inversions, whose details are  
described in Sect.~\ref{structuralinversions}.

\subsection{Forward modelling}
Those reference asteroseismic models are obtained with the local optimisation Levenberg-Marquadt 
algorithm. The algorithm works in connection with our stellar evolution code CLES, and the details of this 
implementation can be found in \citet{miglio05}. The quality of the iterative fits to the observational constraints are 
evaluated via the merit function 

\begin{equation}
\chi^2=\sum_{i=1}^{\mathrm{N}_{\mathrm{obs}}} 
\frac{(X_{\mathrm{obs,i}}-X_{\mathrm{th,i}})^2}{\sigma_i^2}
\label{eq1-section2},
\end{equation}where $X_{\mathrm{obs,i}}$ and $X_{\mathrm{th,i}}$ are respectively the observational constraints and their theoretical 
counterparts from the stellar models. The $\sigma_i^2$ are the observational errors associated with $X_{\mathrm{obs,i}}$. 
The different observations $X_{\mathrm{obs,i}}$, both seismic and 
non-seismic, are summarised in Table~\ref{table-explicite-models}. The minimisation is adapted to model binary 
stars assuming common formation. We hence impose the same initial chemical composition and the same age for both models. 
The number of observational constraints, $\mathrm{N}_{\mathrm{obs}}$, includes  those on both stellar 
components, A and B. 

In total there is a set of seven free parameters to be adjusted in the models: the individual masses, 
M$_{\textrm{A}}$ and M$_{\textrm{B}}$; the mixing-length parameters, $\alpha_{\mathrm{MLT},A}$ and 
$\alpha_{\mathrm{MLT},B}$; the common age;  the common initial chemical composition; X$_0$ (initial mass fraction of 
H) and Z/X$|_0$ (with Z$_0$ initial mass fraction of metals).

The Levenberg-Marquadt approach is a stable algorithm switching continuously between the inverse Hessian method and 
steepest descent method, converging to the local minimum closest to the initial guess \citep[see e.g.][]{bevington03}. 
Thus, inherent to a local approach, we may risk  getting stuck at exploring a sole minimum valley if some of the 
observational constraints are given with high precision. To control the robustness of the solution we did, for 
each given set of observational constraints, several runs of optimisation by varying the initial guess stellar 
parameters for each run. We were careful about the precision error of the constraints; overprecision can prevent the method from exploring 
accurately the parameter space even when varying the initial guess. When this occured in our calibrations, 
we relaxed the dominant constraint by adopting a 3$\sigma$ error instead of 1$\sigma$, and explicitly mention it.

\subsection{Structural inversions}
\label{structuralinversions}

We carry out seismic inversions of structural indicators following \citet{reese12} and \citet{buldgenmeand,buldgenS} to 
provide tighter constraints on the properties of the system. The reference models are those obtained first from forward asteroseismic modelling. The inversions are based on the linear relation between relative frequency differences and relative differences in quantities, such as   adiabatic sound speed, density, or adiabatic exponent derived in \citet{dziembowski90}. These relations can be written as
\begin{align}
\frac{\delta \nu}{\nu} (n,\ell)=\int_{0}^{R}K^{n,\ell}_{s_{1},s_{2}}\frac{\delta s_{1}}{s_{1}}dr + 
\int_{0}^{R}K^{n,\ell}_{s_{2},s_{1}}\frac{\delta s_{2}}{s_{2}}dr + \mathcal{F}_{\mathrm{Surf}}, \label{Eq:FreqStruc}
\end{align}
with $\frac{\delta x}{x}=\frac{x_{\mathrm{Obs}}-x_{\mathrm{Ref}}}{x_{\mathrm{Ref}}}$. Here $x$ can be a frequency ($\nu$) or a model quantity (denoted here as $s_{1}$ or $s_{2}$), such as   the density ($\rho$), the squared 
adiabatic 
sound speed ($c^{2}=\frac{\Gamma_{1}P}{\rho}$ with $P$ the pressure and $\Gamma_{1}=\frac{\partial \ln P}{\partial \ln 
\rho}\vert_{S}$, the adiabatic exponent, with $S$ the entropy); $r$ and $R$ are the distance from centre and radius, 
respectively. The subscript `Obs' denotes quantities of the observed 
target, whereas `Ref' denotes quantities related to the reference model of the inversion, obtained here through 
forward 
modelling. In Eq.~\ref{Eq:FreqStruc} the $K^{n,\ell}_{s_{i},s_{j}}$ functions denote the 
structural kernels associated with the linear integral relations between structure and frequencies. The 
$\mathcal{F}_{\mathrm{Surf}}$ function denotes the  surface effect term (used to model the influence of the 
surface regions where the hypotheses used to derive Eq.~\ref{Eq:FreqStruc} break down) on the frequencies.

In this study two indicators were used for both stars, namely the mean density, $\bar{\rho}$, and the 
$S_{\mathrm{Core}}$ 
indicator from \citet{buldgenS}. The integral definitions of these quantities are
\begin{align}
\bar{\rho}= \int_{0}^{R}4\pi r^{2} \rho dr, \label{Eq:MeanDens} \\
S_{\mathrm{Core}} =  \int_{0}^{R} \frac{f(r)}{S_{5/3}} dr, \label{Eq:IndicSCore}
\end{align}
where $S_{5/3}=\frac{P}{\rho^{5/3}}$ is the entropy proxy and   $f(r)$   is the weight function associated with the 
$S_{\mathrm{Core}}$ indicator: 
\begin{align}
f(r)=& r \left( a_{1} \exp \left(-a_{2}\left( \frac{r}{R}-a_{3} \right)^{2} \right) + a_{4} \exp 
\left(-a_{5}\left(\frac{r}{R}-a_{6} \right)^{2}\right) \right) \nonumber \\
&\tanh \left(a_{7} \left(1-\frac{r}{R} \right) \right). \label{eqRefTarCore}
\end{align}
In this last expression the values of $a_{i}$ are fixed so as to get the best compromise between extracting 
as much information on the core properties and allowing an accurate fit of the target function by the restricted amount 
of frequencies. We used the $(\rho,\Gamma_{1})$ structural kernels for the mean density inversions and the 
$(S_{5/3},Y)$ 
structural kernels, $Y$ being the helium mass fraction, for the $S_{\mathrm{Core}}$ inversion. 

The $S_{\mathrm{Core}}$ indicator is defined in \citet{buldgenS} as an indicator of the presence of 
convective cores in the solar-like oscillators. The physical motivation behind the use of the indicator is that the 
quantity $S_{5/3}=\frac{P}{\rho^{5/3}}$ will present a plateau in adiabatic convective regions. The height of this 
plateau depends crucially on the properties of the convective core. Thus, by inverting the $S_{\mathrm{Core}}$ 
indicator, going as $1/S_{5/3}$, we would actually be able to detected the presence of a convective core by noticing 
significant corrections by the inversion to the indicator value of a given reference model.

The trade-off parameters of the inversion were optimised by testing the inversion between various models in the sample 
of references determined by forward modelling, using the same modes and uncertainties as those of the observed data.

\section{Results of the asteroseismic forward modelling}
\label{section4}

We calibrated series of models for $\alpha$~Cen~A and B by varying both classical and seismic observational 
constraints. We present the results of the different asteroseismic modellings according to the indicators used as 
constraints. The first series of results is based on a $\Delta \nu$--$\delta \nu$ combination. The other set of results 
is constrained with the individual frequency ratios as defined in Eq.~\ref{eq3}. Hereafter the names of the resulting 
models starting with A (resp. B) correspond to calibrations of $\alpha$~Cen~A (resp. $\alpha$~Cen~B).

\subsection{Calibrations based on $\Delta \nu$--$\delta \nu$}

The whole set of calibrations in this section used as asteroseismic constraints the $\Delta \nu$--$\delta \nu$ 
indicators computed following the method described in Sect.~\ref{sectionseismic}. We present them in two categories, 
according to the non-seismic data that were adopted as constraints, which are detailed in 
Table~\ref{table-explicite-models}.  We choosed either a combination based on 
(R$_{\textrm{A}}$,R$_{\textrm{B}}$,T$_{\textrm{A,eff}}$,T$_{\textrm{B,eff}}$) or 
($\kappa$,$\Gamma$,L$_{\textrm{A}}$,L$_{\textrm{B}}$). In the latter case omitting the luminosity could induce a lack 
of information on the evolutionary stage of the stars. Each lowest oscillation frequency of the two stars were 
also used to guide the calibration process and avoid degeneracy linked to iso-frequency separation solutions.

The input physics was varied to test the effects of different opacity dataset and chemical mixtures. The different inputs 
are summarised in Table~\ref{table-explicite-models-2}. Some calibrations also included core overshooting, with 
$\alpha_{\textrm{ov}}=0.10$ or $0.20$. The parameters of the models resulting of the calibrations are presented in 
Table~\ref{table-results}\footnote{Due to the assumption of common formation, the parameters that $\alpha$~Cen~B has 
in common with those of $\alpha$~Cen~A are not repeated.}.

\begin{table*}
\caption{Stellar model parameters of the various asteroseismic forward modellings}             
\label{table-results}      
\centering          
\begin{tabular}{l c c c c c c c c c c c c}     
\hline\hline       
Model & M & $\Delta $M & $\alpha_{\textrm{MLT}}$ & $\Delta \alpha_{\textrm{MLT}}$ & X$_0$ & 
$\Delta$X$_0$ & Z/X|$_0$ & $\Delta$Z/X|$_0$ & Z/X|$_\textrm{S}$ & age & $\Delta$ age & $\chi^2$\\ 
 & [M$_{\odot}$] & [M$_{\odot}$] & & & & & & & & [Gyr] & [Gyr] 
& \\
\hline
\hline

    A-$\Delta_{\nu}\delta_{\nu}$-R,T$_{\textrm{eff}}$  &    1.105  &    0.012  &     2.07  &     0.13  &    0.699  &    
0.010  &   0.0373  &   0.0018  &   0.0298  &     6.97  &     1.13  &    13.32  \\
    B-$\Delta_{\nu}\delta_{\nu}$-R,T$_{\textrm{eff}}$  &    0.919  &    0.009  &     2.09  &     0.18  &       --  &    
 
  --  &       --  &       --  &   0.0323  &       --  &       --  &       --  \\ \hline
    A-$\Delta_{\nu}\delta_{\nu}$-M$_{\rm frac}$,R$_{\rm ratio}$  &    1.112  &    0.014  &     2.22  &     0.06  &    
0.711  &   
 0.010 
 &   0.0402  &   0.0019  &   0.0321  &     8.56  &     0.39  &    69.72  \\
    B-$\Delta_{\nu}\delta_{\nu}$-M$_{\rm frac}$,R$_{\rm ratio}$  &    0.940  &    0.012  &     2.43  &     0.08  &       
--  &   
    -- 
 &       --  &       --  &   0.0342  &       --  &       --  &       --  \\ \hline
       A-$\Delta_{\nu}\delta_{\nu}$-M$_{\rm frac}$,R$_{\rm ratio}$-iniLM  &    1.113  &    0.011  &     2.25  &     0.07 
 &    
0.710  & 
   0.009  &   0.0405  &   0.0018  &   0.0323  &     8.58  &     0.41  &    61.41  \\
    B-$\Delta_{\nu}\delta_{\nu}$-M$_{\rm frac}$,R$_{\rm ratio}$-iniLM  &    0.941  &    0.009  &     2.44  &     0.08  & 
      
--  &    
   --  &       --  &       --  &   0.0344  &       --  &       --  &       --  \\ \hline
   A-$\Delta_{\nu}\delta_{\nu}$-M$_{\rm frac}$,R$_{\rm ratio}$-OPLIB  &    1.097  &    0.018  &     2.21  &     0.08  &  
  0.711 
 &    
0.011  &   0.0388  &   0.0018  &   0.0310  &     8.37  &     0.29  &    151.62  \\
    B-$\Delta_{\nu}\delta_{\nu}$-M$_{\rm frac}$,R$_{\rm ratio}$-OPLIB  &    0.927  &    0.015  &     2.48  &     0.08  & 
      
--  &    
   --  &       --  &       --  &   0.0330  &       --  &       --  &       --  \\ \hline
   A-$\Delta_{\nu}\delta_{\nu}$-M$_{\rm frac}$,R$_{\rm ratio}$-GN93  &        1.119  &    0.013  &     2.21  &     0.07  
&    
0.705  &  
  0.009  &   0.0512  &   0.0025  &   0.0411  &     8.55  &     0.37  &    59.74  \\
    B-$\Delta_{\nu}\delta_{\nu}$-M$_{\rm frac}$,R$_{\rm ratio}$-GN93  &       0.946  &    0.011  &     2.35  &     0.07  
&       
--  &  
     --  &       --  &       --  &   0.0436  &       --  &       --  &       --  \\ \hline
    A-$\Delta_{\nu}\delta_{\nu}$-M$_{\rm frac}$,R$_{\rm ratio}$-GN93-Ov  &    1.118  &    0.021  &     1.97  &     0.08  
&    
0.700  &  
  0.015  &   0.0472  &   0.0024  &   0.0381  &     7.20  &     0.43  &   113.99  \\
    B-$\Delta_{\nu}\delta_{\nu}$-M$_{\rm frac}$,R$_{\rm ratio}$-GN93-Ov  &    0.943  &    0.018  &     2.24  &     0.09  
&       
--  &  
     --  &       --  &       --  &   0.0409  &       --  &       --  &       --  \\ \hline
       A-$\Delta_{\nu}\delta_{\nu}$-M$_{\rm frac}$,R$_{\rm ratio}$-Ov  &    1.100  &    0.019  &     2.17  &     0.10  & 
   
0.718  &    
0.012  &   0.0343  &   0.0018  &   0.0276  &     8.66  &     0.44  &    93.73  \\
    B-$\Delta_{\nu}\delta_{\nu}$-M$_{\rm frac}$,R$_{\rm ratio}$-Ov  &    0.929  &    0.016  &     2.39  &     0.05  &    
   --  
&       
--  &       --  &       --  &   0.0291  &       --  &       --  &       --  \\ \hline
   A-$\Delta_{\nu}\delta_{\nu}$-M$_{\rm frac}$,R$_{\rm ratio}$-Ov0.20 &    1.119  &    0.025  &     1.86  &     0.08  &  
  0.705 
 &    
0.013  &   0.0368  &   0.0017  &   
0.0294  &     6.64  &     0.39  &   148.38  \\
   B-$\Delta_{\nu}\delta_{\nu}$-M$_{\rm frac}$,R$_{\rm ratio}$-Ov0.20 &    0.943  &    0.022  &     2.28  &     0.04  &  
     -- 
 &     
  --  &       --  &       --  &   
0.0321  &       --  &       --  &       --  \\ \hline

       A-$\Delta_{\nu}\delta_{\nu}$-R,T$_{\textrm{eff}}$-Surf  &    1.088  &    0.011  &     1.84  &     0.04  &    
0.689  
&    0.008  &   0.0367  &   0.0017  &   0.0293  &     6.19  &     0.61  &    24.03  \\
    B-$\Delta_{\nu}\delta_{\nu}$-R,T$_{\textrm{eff}}$-Surf  &    0.911  &    0.007  &     2.01  &     0.02  &       --  
&   
    --  &       --  &       --  &   0.0320  &       --  &       --  &       --  \\ \hline
    
   A-$\Delta_{\nu}\delta_{\nu}$-M$_{\rm frac}$,R$_{\rm ratio}$-Surf  &    1.049  &    0.008  &     2.07  &     0.09  &   
 0.699  
& 
   0.008  &   0.0319  &   0.0018  &   0.0250  &     8.12  &     0.33  &    80.94  \\
   B-$\Delta_{\nu}\delta_{\nu}$-M$_{\rm frac}$,R$_{\rm ratio}$-Surf  &    0.886  &    0.006  &     2.05  &     0.02  &   
    --  
& 
      --  &       --  &       --  &   0.0269  &       --  &       --  &       --  \\ \hline
      
   A-$\Delta_{\nu}\delta_{\nu}$-R,T$_{\textrm{eff}}$-Z/X$_{\textrm{BTK08}}$  &    1.117  &    0.011  &     1.95  &     
0.13  &    0.713  &    
0.008  &   0.0318  &   0.0015  &   0.0255  &     6.24  &     0.65  &    39.09  \\
   B-$\Delta_{\nu}\delta_{\nu}$-R,T$_{\textrm{eff}}$-Z/X$_{\textrm{BTK08}}$  &    0.924  &    0.009  &     2.00  &     
0.12  &       --  &      
 --  &       --  &       --  &   0.0279  &       --  &       --  &       --  \\ \hline
 
   AA-$\Delta_{\nu}\delta_{\nu}$-M$_{\rm frac}$,R$_{\rm ratio}$-Z/X$_{\textrm{BTK08}}$  &    1.109  &    0.027  &     
1.94  &    
 
0.08  &    0.719  &    0.015  &   0.0330  &   0.0015  &   0.0263  &     7.28  &     0.38  &    93.31  \\
   BB-$\Delta_{\nu}\delta_{\nu}$-M$_{\rm frac}$,R$_{\rm ratio}$-Z/X$_{\textrm{BTK08}}$  &    0.936  &    0.023  &     
2.23  &    
 
0.05  &       --  &       --  &       --  &       --  &   0.0285  &       --  &       --  &       --  \\ \hline

   A-r$_{10}$-M,R  &    1.112  &    0.009  &     1.88  &     0.22  
&    0.701  &    0.009  &   0.0364  &   0.0018  &   0.0292  &     6.25  &     1.40  &    10.71  \\
   B-r$_{10}$-M,R &    0.927  &    0.006  &     2.05  &     0.16  & 
      --  &       --  &       --  &       --  &   0.0318  &       --  &       --  &       --  \\ \hline
      
   A-r$_{10}$-M,R-Ov  &    1.115  &    0.008  &     1.78  &     0.14  &    0.698  &    0.009  &   0.0355  &   0.0017  & 
  
0.0285  &     5.55  &     1.04  &    28.40  \\
   B-r$_{10}$-M,R-Ov  &    0.925  &    0.005  &     1.96  &     0.13  &       --  &       --  &       --  &       --  & 
  
0.0314  &       --  &       --  &       --  \\ \hline

   A-r$_{02}$-M,R   &    1.115  &    0.006  &     1.93  &     0.04  &    0.696  &    0.006  &   0.0388  &   0.0018  &  
 0.0312  &     6.35  &     0.25  &    26.73  \\
   B-r$_{02}$-M,R &    0.930  &    0.005  &     2.09  &     0.05  &       --  &       --  &       --  &       --  &   
0.0340  &       --  &       --  &       --  \\ \hline

   A-r$_{02}$-M,R-Ov &    1.114  &    0.007  &     1.87  &     0.04  &    0.701  &    0.008  &   0.0362  &   0.0017  &  
 
0.0291  &     6.20  &     0.26  &    27.03  \\
   B-r$_{02}$-M,R-Ov &    0.926  &    0.005  &     2.04  &     0.03  &       --  &       --  &       --  &       --  &  
 
0.0318  &       --  &       --  &       --  \\ \hline

   A-r$_{13}$-M,R  &    1.115  &    0.007  &     1.84  &     0.03  &    0.700  &    0.007  &   0.0362  &   0.0018  &  
 0.0291  &     5.93  &     0.21  &    18.79  \\
   B-r$_{13}$-M,R  &    0.928  &    0.005  &     2.02  &     0.04  &       --  &       --  &       --  &       --  &  
 0.0319  &       --  &       --  &       --  \\ \hline
 
   A-r$_{13}$-M,R-Ov &    1.115  &    0.007  &     1.84  &     0.04  &    0.700  &    0.007  &   0.0360  &   0.0017  &  
 
0.0289  &     5.97  &     0.21  &    19.15  \\
   B-r$_{13}$-M,R-Ov &    0.927  &    0.005  &     2.02  &     0.03  &       --  &       --  &       --  &       --  &  
 
0.0317  &       --  &       --  &       --  \\

\hline
\hline                  
\end{tabular}

\end{table*}

\subsubsection{Impact of the non-seismic constraints}
\label{subconstr}
\begin{figure*}[!]

\includegraphics[width=20cm]{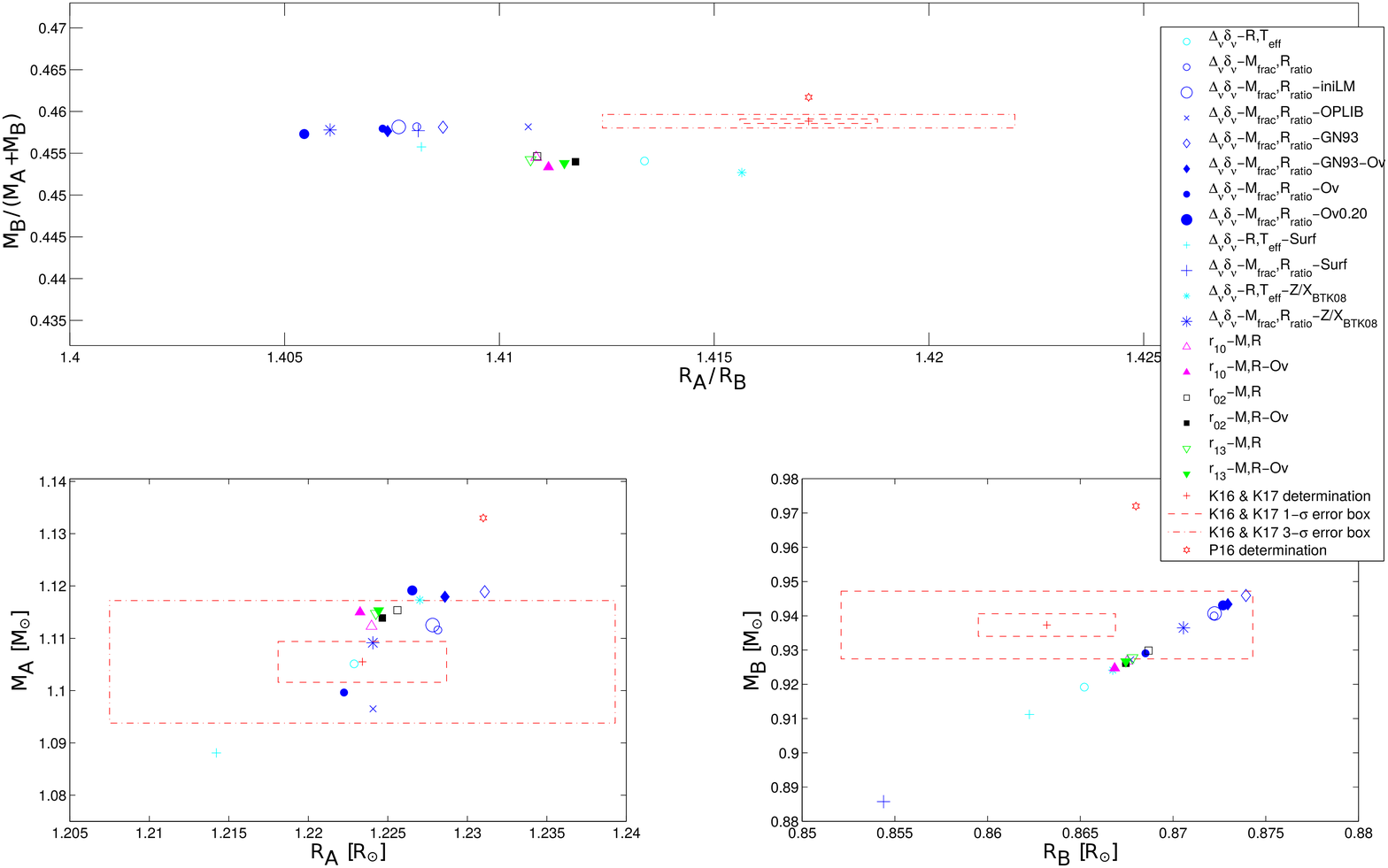}

\caption{Top panel: Inferred fractional mass to ratio of radii  for the $\alpha$~Cen system, obtained through 
various asteroseismic calibrations (see details in Tables~\ref{table-explicite-models} and 
\ref{table-explicite-models-2}). The calibrations are represented by different symbols (see legend). Cyan 
and blue symbols correspond to calibrations made with the $\Delta \nu~-~\langle\delta \nu_0\rangle$ combination as 
seismic 
constraints:  cyan for mass and effective temperature as non-seismic constraints, blue for fractional mass and radius ratio instead. Magenta, black, and green symbols respectively correspond to calibrations with 
$r_{10}$, $r_{02}$, and $r_{13}$ used as seismic constraints. Open symbols indicates that no overshooting was included 
in 
the stellar models, while filled symbols denotes cases where models are computed with 
overshooting. The red cross and the dashed and dot-dashed lines respectively give the values 
inferred from the resolution of the orbital motion by K16 and K17, and the limits of the 1$\sigma$ and 3$\sigma$ error 
box 
on the K16 and K17 determinations. The red star indicates the parameters obtained from the P16 orbital solution. \newline 
Bottom left panel: Same as top panel, but for the inferred mass and radius of $\alpha$~Cen~A. \newline 
Bottom 
right panel: Same as top panel, but for the inferred mass and radius of $\alpha$~Cen~B.}
\label{fig-rad-mas}
\end{figure*}

We show in Fig.~\ref{fig-rad-mas} the inferences on the masses and radii. In the top panel are presented results 
expressed as the fractional mass and the ratio of radii of the system. They can be analysed following two subsets, depending on the 
non-seismic constraints used.

\begin{figure*}[t!]
                \includegraphics[scale=0.45]{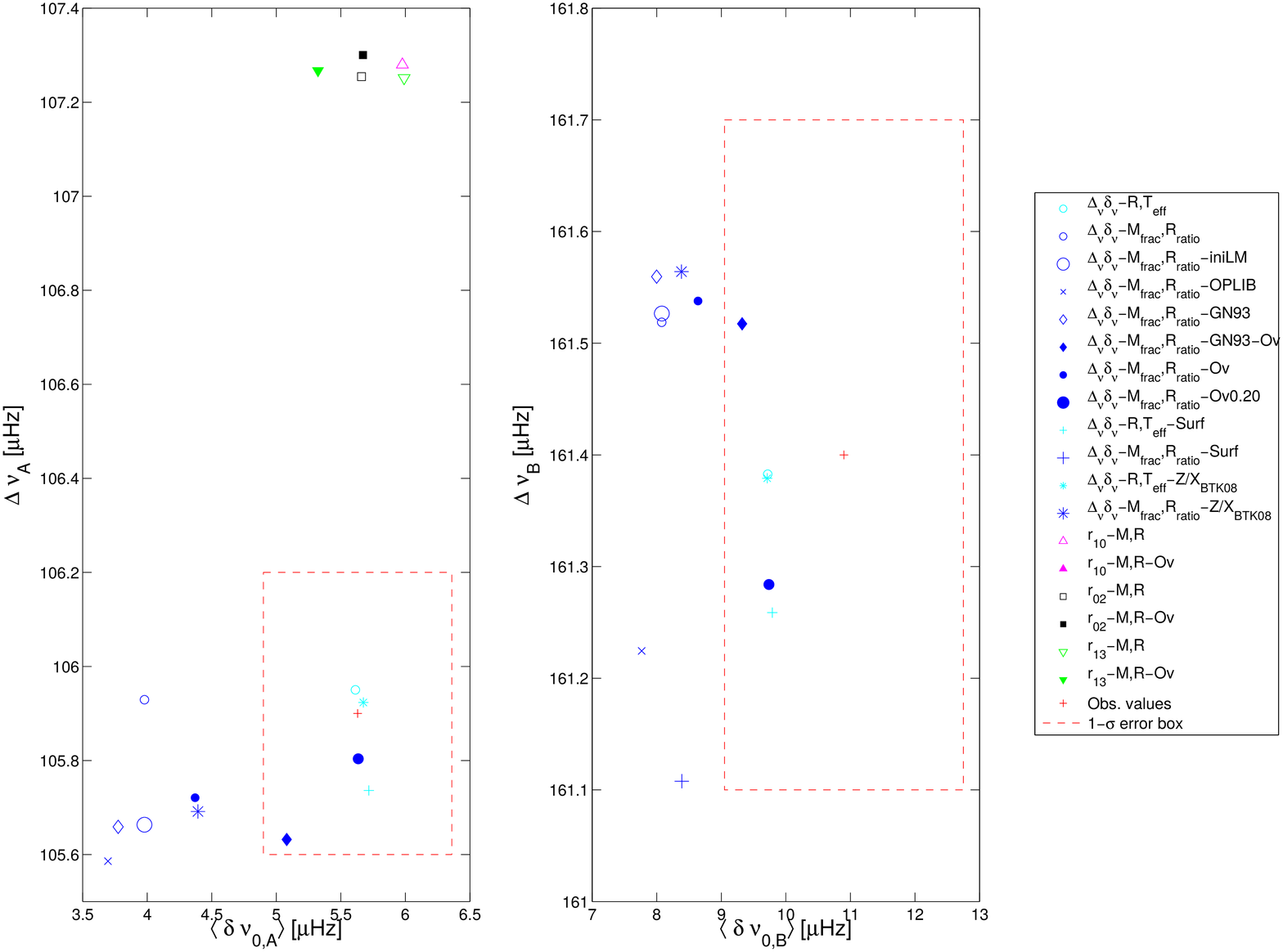}
                \hspace{2.5cm} \includegraphics[scale=0.35]{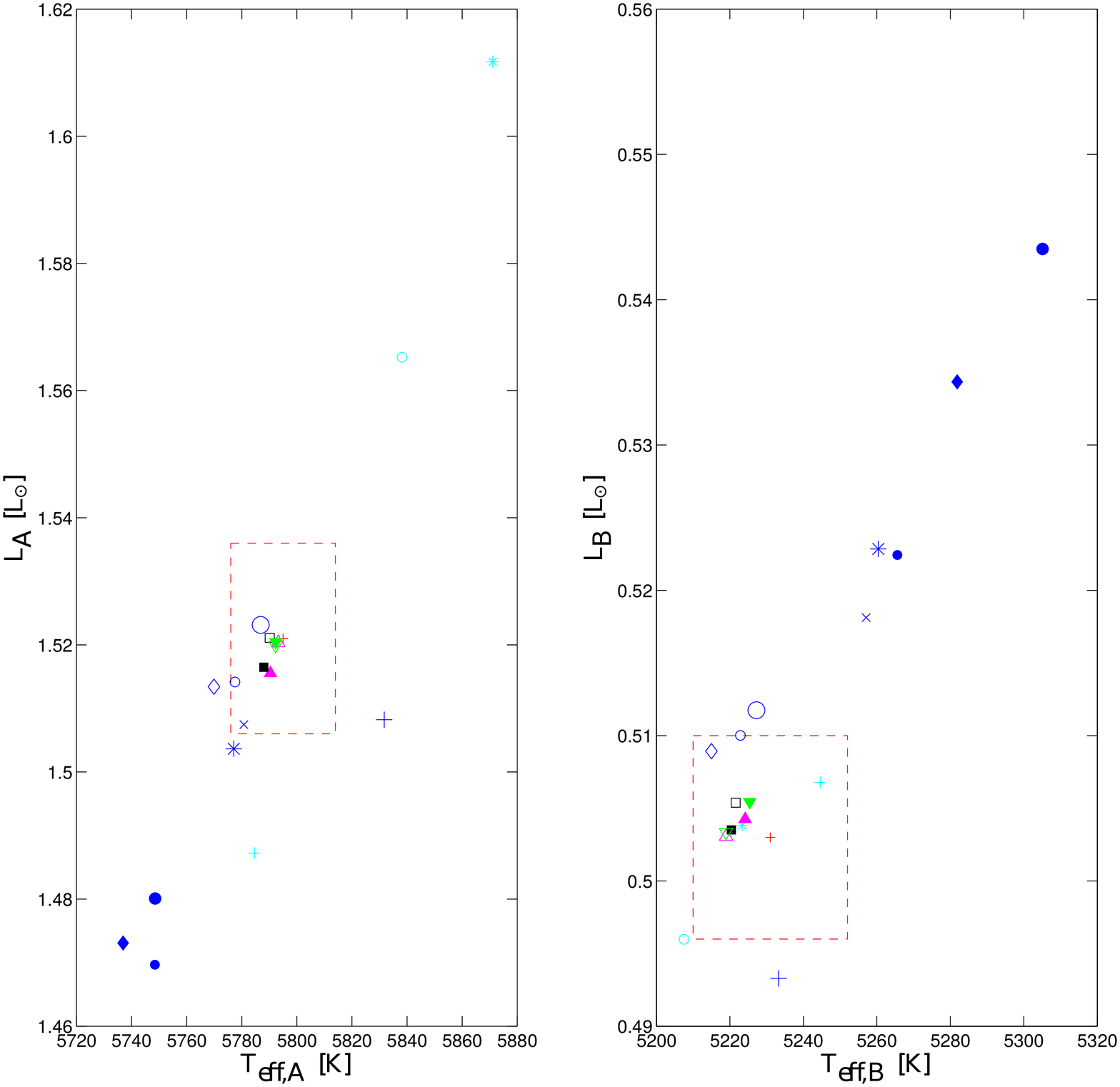}
                
        \caption{Top left and right panels: Large and small frequency separations of the final 
inferred models of $\alpha$~Cen~A and B, for the various asteroseismic calibrations. Symbols are the same as  those used 
in Fig.~\ref{fig-rad-mas}, except for the red cross  and dashed line, which represent respectively the observed 
values and their 1$\sigma$ error box, that we derived from the pulsation frequency analysis made by \citet{demeulenaer10}. Bottom panels:  Hertzsprung-Russell diagram for the results of the different calibrations  compared 
 to the observational values of K17. The A and B components are depicted respectively in the left and 
right panels. The legend details are the same as in the top panels. }
                \label{fig-dnu}
\end{figure*}

The first subset covers the calibrations in which the fractional mass and radius ratios were used as constraints. The 
names of the calibration include `M$_{\rm frac}$,R$_{\rm ratio}$' and are depicted in blue. These models are 
the 
only ones to fall within or close to the 3$\sigma$ error boxes on the fractional mass derived by K16. The radius ratio 
$\Gamma$ is systematically lower than the K17 ratio: all these models clearly fail to reproduce this constraint. The 
three 
models with overshooting (and the only models in this subset with a convective core) and the model with a lower 
metallicity reproduce $\Gamma$ with the lowest accuracy. Their position in a Hertzsprung-Russell diagram in the bottom 
panel of
Fig.~\ref{fig-dnu} confirms this tendency, as the four models differ by more than 1$\sigma$   from the
luminosities of $\alpha$~Cen~A and $\alpha$~Cen~B.

Results for $\alpha$~Cen~A (in the bottom left panel of Fig.~\ref{fig-rad-mas}) indicate that only the 
lower metallicity model predicts a mass within 1$\sigma$ to the mass of K16. The models with the 
GN93 
mixture tend to masses higher than that of K16, and closer to that of P16. All the models but one predict a radius larger than K17, 
yet well within the 1$\sigma$ limit on it.  

Expected from the offset in the fitting of $\Gamma$, the radii predicted for $\alpha$~Cen~B (bottom right panel of
Fig.~\ref{fig-rad-mas}) by this subset of models present a systematic shift (but within 3$\sigma$) to the K17 
value.  The predicted masses are all within 1$\sigma$ or 3$\sigma$ of that of K16, except the model including surface 
effect correction. This model has a mass significantly lower than the K16 and P16 values. Again, the mass of the models 
calibrated with the GN93 mixture are the highest.

In Fig.~\ref{fig-dnu} all the models of the B component (excepting the one with surface effect correction) 
reproduce within 1$\sigma$ its large separation and are close to the lower 1$\sigma$ limit for the small separation. 
As previously mentioned, this subset of models suggests higher masses and larger radii for the B component than the K16 values. The 
good fitting of the large separation (and thus the mean density) could either reveal a discrepancy between the seismic and 
astrometric plus interferometric solutions, or a degeneracy in the seismic solution. Given 
the difference between the observed T$_{\textrm{B,eff}}$ and especially L$_{\textrm{B}}$ (bottom right panel of 
Fig.~\ref{fig-dnu}) and those of the models, it is likely that we converged to a degenerate solution where the fitting 
of the large separation was privileged.

On the contrary, for the A component (top left panel of Fig.~\ref{fig-dnu}) the fit of the large separation 
systematically tends to larger values, hence overestimating its mean density. This same set of models reproduces the 
R$_{\textrm{A}}$ value from K16. Since they fit  L$_{\textrm{A}}$, the same is true for T$_{\textrm{A,eff}}$,  
which is correlated to the determination of R$_{\textrm{A}}$. Since M$_{\textrm{A}}$ predicted by the seismic models 
are higher than the K16 value, it likely reveals the origin of the mean density overestimation (which 
is confirmed by the inversions in Sect.~\ref{section-inv-seismo}). We note that due to the high precision on $\kappa$ and 
$\Gamma$ from the latter solution, we deal with a delicate trade-off in the adjustments of the seismic and non-seismic 
contributions in the $\chi^2$.

\begin{figure*}[t!]
        
                \includegraphics[scale=0.55]{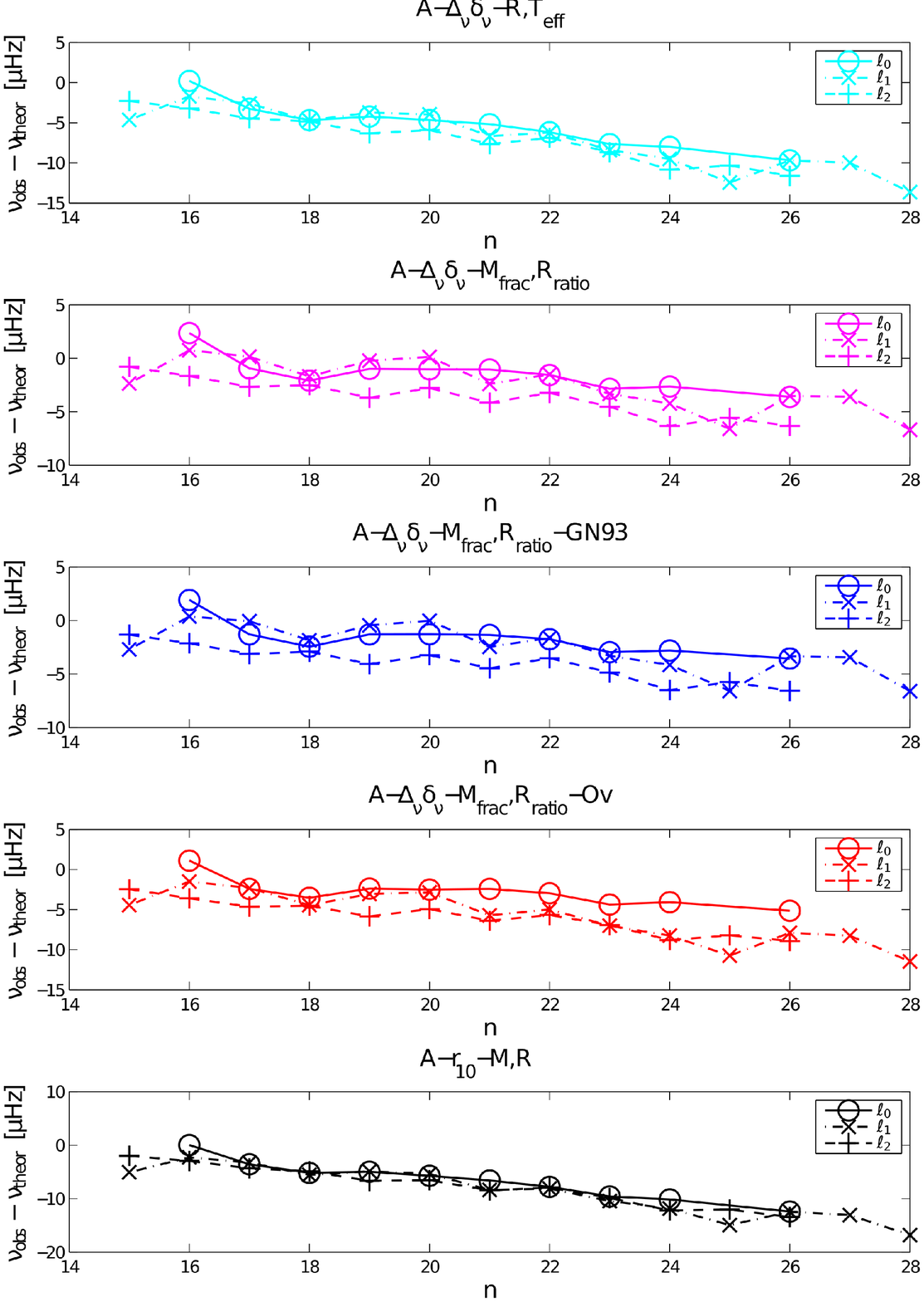}

        \caption{Differences between the $\alpha$~Cen~A observed frequencies and the 
theoretical frequencies from the models of the following calibrations (from top to bottom): A-$\Delta_{\nu}\delta_{\nu}$-R,T$_{\textrm{eff}}$; 
A-$\Delta_{\nu}\delta_{\nu}$-M$_{\rm frac}$,R$_{\rm ratio}$; A-$\Delta_{\nu}\delta_{\nu}$-M$_{\rm frac}$,R$_{\rm 
ratio}$-GN93; 
A-$\Delta_{\nu}\delta_{\nu}$-M$_{\rm frac}$,R$_{\rm ratio}$-Ov; and A-r$_{10}$-M,R. The different $\ell$ degrees of the 
modes 
shown are indicated in the insets. }
                \label{fig-diff-freqA}
\end{figure*}

We also looked at the differences between observed and theoretical frequencies, which are shown in 
Fig.~\ref{fig-diff-freqA} for $\alpha$~Cen~A. In the three panels we see  for this subset of models that adopting the 
AGSS09 or GN93 mixture does not significantly change the precision of the frequency fitting. We note that for a model 
with overshooting (and the presence of a convective core) the differences with the observed frequencies are amplified.

The  second subset of solutions includes the three calibrations that are directly based on individual masses and effective 
temperatures (`R,T$_{\textrm{eff}}$' in their names, depicted in cyan). They result in significantly lower $\kappa$ values than in K16 and P16. They predict individual masses lower by  
$\sim0.02-0.03$M$_{\odot}$ than the K16 value for the B component, and close to the K16 value for the A component (see 
bottom panels in Fig.~\ref{fig-rad-mas}). This naturally results in decreasing the fractional mass $\kappa$ in 
comparison to K16. Except for the calibration with the surface effect correction, the two other 
calibrations reproduce the individual radii of both components, and hence the radius ratio $\Gamma$. All three 
calibrations fit the $\Delta\nu~-~\langle\delta \nu_0\rangle$ values for the B component within 1$\sigma$. They also fit 
 the $\langle\delta \nu_0\rangle$ of $\alpha$~Cen~A (see Fig.~\ref{fig-dnu}). However, as in every other case, the fit 
of $\Delta\nu_{\textrm{A}}$ results in larger values. For this smaller subset of models, L$_{\textrm{A}}$ is 
also poorly reproduced, which again raises the question of a discrepancy on the stellar constraints between seismic and 
non-seismic observables.

If we look at the quality of the fit based on the individual seismic frequencies in Fig.~\ref{fig-diff-freqA}, we do 
not see any significant impact depending on the choice of the non-seismic constraint (see top two   panels). More 
interestingly, we note a clear increasing trend with radial order in the differences between observed and theoretical 
frequencies, as  in the \citet{miglio05} and \citet{eggenberger04} studies. The amplitudes we get, of a 
few $\mu$Hz, are of the same order as in the asteroseismic modelling by \citet[their Fig.~6]{miglio05} and lower than in \citet{eggenberger04}.

The choice of the non-seismic constraints essentially impacts the masses and radii determinations of the system, as expected. Other inferred parameters do not show a particular correlation depending of the choice of 
non-seismic observables. In particular, a similar range of ages is predicted by the two subsets, 
between $\sim$6.2 and $\sim$8.6~Gyr. This is in line with the $\langle\delta \nu_0\rangle$ values (a marker of 
the 
evolution), which are similar for the two subsets.

\subsubsection{GN93 vs AGSS09: Impact on the mass determination}
\label{gn93vsagss09}

We adopted the surface metallicity of \citet{portodemello08} as constraint for $\alpha$~Cen~A and B in combination 
with the AGSS09 chemical mixture and OPAL opacities as input of the models in the majority of our calibrations. The means of the masses derived in these calibrations give $\langle$M$_{\textrm{A}}\rangle=1.098 \pm 
0.014$ M$_{\odot}$ and $\langle$M$_{\textrm{B}}\rangle=0.924\pm0.011$ M$_{\odot}$. On the other hand, the weighted 
means of the two 
calibrations adopting the GN93 mixture yield $\langle$M$_{\textrm{A}}\rangle=1.118 \pm 0.017$ M$_{\odot}$ and 
$\langle$M$_{\textrm{B}}\rangle=0.945\pm0.015$ M$_{\odot}$. The reason for the decrease in mass is likely to 
be related 
with a decrease in the opacity in the models with AGSS09. All things equal, for a given luminosity a 
metal-poor model will be less massive than a more metal-rich counterpart. However, because of degeneracies between the 
free parameters of the fitting, other elements, such as the initial composition, may generate differences in the output 
of calibrations with various physical ingredients. Hence we cannot exclude a combination of effects affecting the 
inferred masses when changing the chemical mixture. 

For comparison with the literature, in their study \citet{miglio05} did not observe a decrease in mass when switching 
from GN93 to the then metallicity downward-revised \citet{asplund05} solar mixture. However, \citet{miglio05} did only 
one study case based on these more metal-poor abundances (the A5 and B5 calibration in their Table 2), and employed a 
different seismic indicator. If we look at the results with the GN93 mixture, our values for 
$\langle$M$_{\textrm{A}}\rangle$ and 
$\langle$M$_{\textrm{B}}\rangle$ generally exceed  by $\sim$0.015~M$_{\odot}$ those derived by \citet{miglio05}. Other 
asteroseismic studies \citep[e.g.][]{bazot12, nsamba18} present methods and physics assumptions that  diverge from  ours, which hamper comparisons at a similar level of detail. 

If we now look at the masses derived from the orbital solutions by K16 and P16, the update of the chemical mixture in 
the stellar models with AGSS09 predict a lower asteroseismic primary mass, but distant by less than its 1$\sigma$ 
error from  that of K16 (M$_{\textrm{A}}=1.1055 \pm 0.0039$ M$_{\odot}$). It is also lower than P16 
(M$_{\textrm{A}}=1.133 \pm 0.005$ M$_{\odot}$), but distant by more than 2$\sigma$.  For the secondary star, 
the asteroseismic mass (still considering AGSS09) is in agreement almost by 1$\sigma$ with that of K16 
(M$_{\textrm{B}}=0.9373 \pm 0.0033$ M$_{\odot}$), but is in disagreement with P16 (M$_{\textrm{B}}=0.972 \pm 0.005$ 
M$_{\odot}$). This asteroseismic determination is hence in overall agreement with the K16 solution, not with the P16 
solution.

To understand the source of this disagreement, we first recall that the individual masses of K16 and P16 are obtained with 
help of the fractional mass of the system and a determination of the parallax. Since $\kappa$ in K16 and P16 are so 
close, the differences between their estimation of the individual masses arise mainly from a difference in the 
parallax. 
If we compute $\kappa$ with help of $\langle$M$_{\textrm{A}}\rangle$ and $\langle$M$_{\textrm{B}}\rangle$, we 
obtain $\langle\kappa\rangle=0.457~\pm~0.011$, in good agreement with the orbital solutions of K16 and P16. So, two 
sources could explain our disagreement 
with the determination of P16. First, if we consider that our asteroseismic determination of the masses with AGSS09 are 
exact, 
it would indicate an error of accuracy on the parallax adopted in P16. To the contrary, if we consider the solution of P16 to be 
correct, it would question the role and adequacy of using the AGSS09 abundances for stellar models of $\alpha$~Cen~AB 
stars \citep[which are not exactly solar scaled, as shown in e.g.][]{morel18}.

Considering the results with the more metal-rich GN93 mixture, the asteroseismic masses present values  between the  
K16 and P16 solutions. For each component the determination with GN93 is closer to the K16 value 
($\sim 1\sigma$) than to the P16 value ($\sim 2\sigma$).

\subsubsection{OPAL vs OPLIB opacity}

The role of the opacity in the stellar parameter inferences was explored in a calibration made with OPLIB data
($\Delta_{\nu}\delta_{\nu}$-M$_{\rm frac}$,R$_{\rm ratio}$-OPLIB). In comparison to its counterpart calibrations made 
with OPAL opacity ($\Delta_{\nu}\delta_{\nu}$-M$_{\rm frac}$,R$_{\rm ratio}$ and $\Delta_{\nu}\delta_{\nu}$-M$_{\rm frac}$,R$_{\rm 
ratio}$-iniLM), the OPLIB calibration leads to a slight decrease in the masses of the system. This is   expected since the OPLIB opacities are slightly 
lower than OPAL in the radiative regions of solar-type stars \citep{colgan16}.

\subsubsection{Surface metallicity}

The two calibrations where the surface metallicity of \citet{bigot08} is adopted lead to a less clear-cut result. Following the above reasoning on the opacity (see Sect.~\ref{gn93vsagss09}) we 
would expect  to find lower masses 
than in the cases with the higher metallicity of \citet{portodemello08}. However, the masses do not show such a 
systematic trend, and the calibrations with \citet{bigot08} predict similar ages to those with 
\citet{portodemello08}. This actually indicates that the initial abundances adapt to compensate the effect of 
microscopic diffusion that will act on a similar timescale. The models calibrated with \citet{bigot08} thus present 
larger initial hydrogen abundance and lower initial metallicities, which result in a higher opacity in central region 
where free-free light absorption processes dominate. The larger X$_0$ somehow mitigates the effect of a decrease in the 
metallicity and hence it barely affects the resulting masses.

\subsubsection{Surface correction to frequencies}

We included surface effect corrections, following \citet{sonoi15}, in 
the $\Delta_{\nu}\delta_{\nu}$-R,T$_{\textrm{eff}}$-Surf and $\Delta_{\nu}\delta_{\nu}$-M$_{\rm frac}$,R$_{\rm 
ratio}$-Surf 
cases. In the first calibration we see a decrease of $\sim0.01~M_{\odot}$ and $\sim800$~Myr, although of the 
typical order of the error on ages, in the inferred stellar parameters by including surface correction. In the second 
case the masses of each component are the lowest we obtain in the whole set of calibrations. But we find it difficult 
in that case to reproduce the seismic indicators, which are the only ones out the 1$\sigma$ box on $\Delta 
\nu$~-~$\langle\delta \nu_0\rangle$ observed for each component (see large blue crosses   in the top panels of 
Fig.~\ref{fig-dnu}). The surface 
metallicity is also poorly reproduced by this calibration. The corrections actually mostly deteriorate the 
ability to reproduce the $\Delta \nu$ observed value by the two calibrations. The \citet{sonoi15} prescription tends to 
overcorrect even the low frequencies, while the surface effects are generally expected to be predominant at 
higher frequencies. The corrections could thus be overestimated for given frequencies and so introduce a bias in the 
$\Delta \nu$ computed from the stellar models. 

\subsection{Calibrations based on r$_{10}$, r$_{02}$, r$_{13}$}

The calibrations based on frequency ratios (see Eqs.~\ref{eq3}) are essentially sensitive to the central stellar 
conditions. They then suffer a lack of constraints on the global stellar parameters hindering convergence of the 
optimisation process. We  adopted more stringent non-seismic constraints on the global stellar parameters,  namely the K16 individual masses and the 
K17 radii and luminosities. As with the other set of seismic 
constraints, we also used the lowest oscillation frequency of each stellar component to improve the convergence to an 
acceptable seismic solution.

The Levenberg-Marquadt method converges, but results in inferences on the mass and radius for each star 
restricted to a similar narrow range of values, whatever  ratios are used as seismic indicators; for instance, the results reproduce the   R$_{\textrm{A}}$ values  of K17 precisely, and the R$_{\textrm{B}}$ values within
1$\sigma$. As these 
models converge to heavier masses for the A component, and lighter masses for the B component, the $\kappa$ they 
predict has thus an offset with the observed $\kappa$ (see top panel of Fig.~\ref{fig-rad-mas}). 

In Fig.~\ref{fig-dnu}, they all reproduce within 1$\sigma$ the small frequency separations, as expected from their 
sensitivity to stratification in central stellar layers, which is also the case of the $\langle\delta \nu_0\rangle$ 
indicator. 
However, all the results overestimate the large separation for the $\alpha$~Cen~A, while they fit it with good 
accuracy for $\alpha$~Cen~B. This could be expected from the bottom panel of  Fig.~\ref{fig-diff-freqA} where 
the frequencies for a model of the A component calibrated with r$_{10}$ show a larger departure from the observed 
frequencies than when using the $\langle\delta \nu_0\rangle$ as a constraint.

All these calibrations predict ages between $\sim$5.55~and~$\sim$6.35~Gyr, which are younger or at the lower 
limit of the range of ages predicted by the series of models from Sect.~\ref{subconstr}. The model obtained 
with the frequency ratios are an interesting test of the potential for inversions to discriminate age and whether the 
method can improve the accuracy on the ages determined with asteroseismology.

\subsection{Age of the system and presence of a convective core}

\begin{figure*}[!]
        \centering
                \includegraphics[width=8cm]{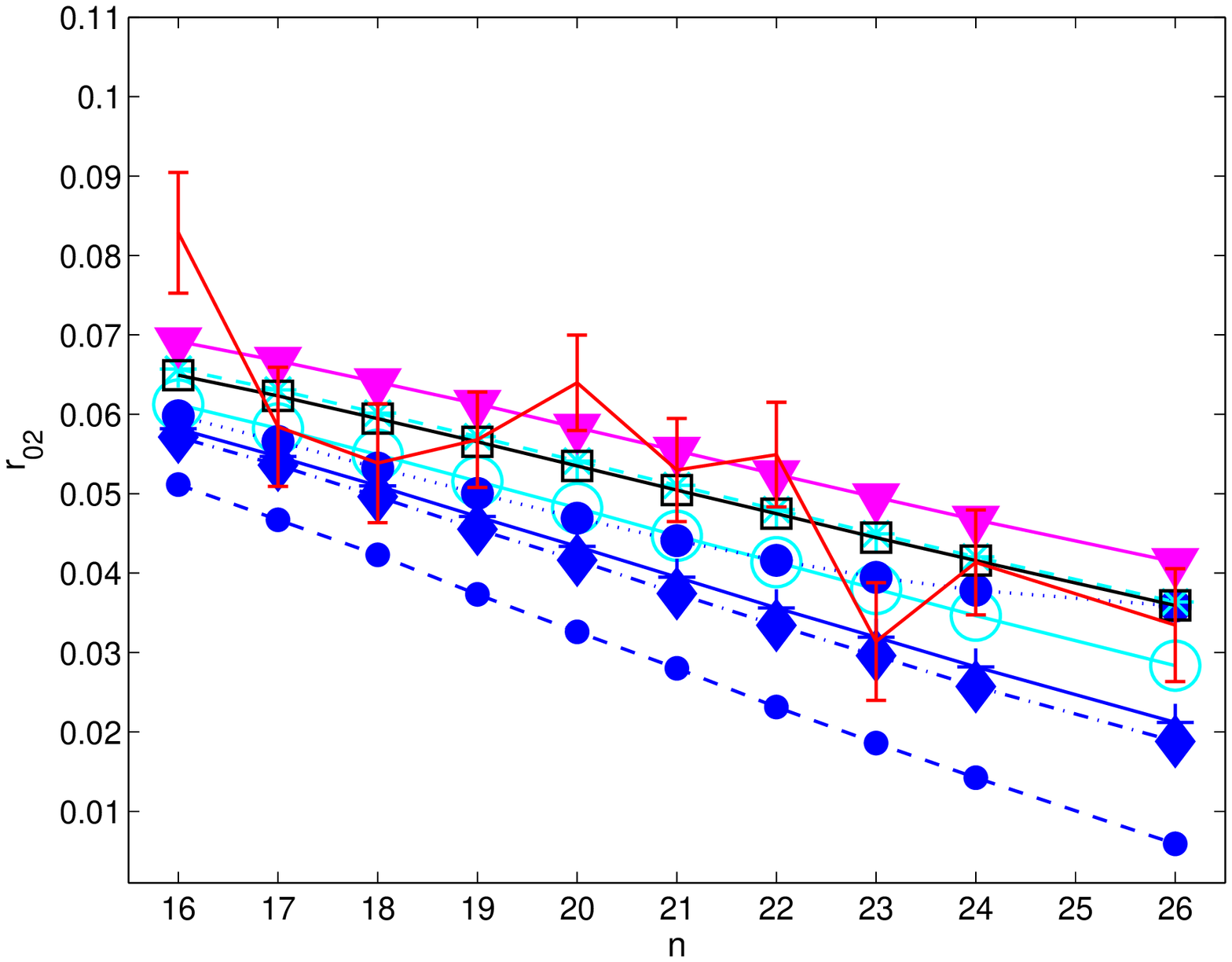}\includegraphics[width=8cm]{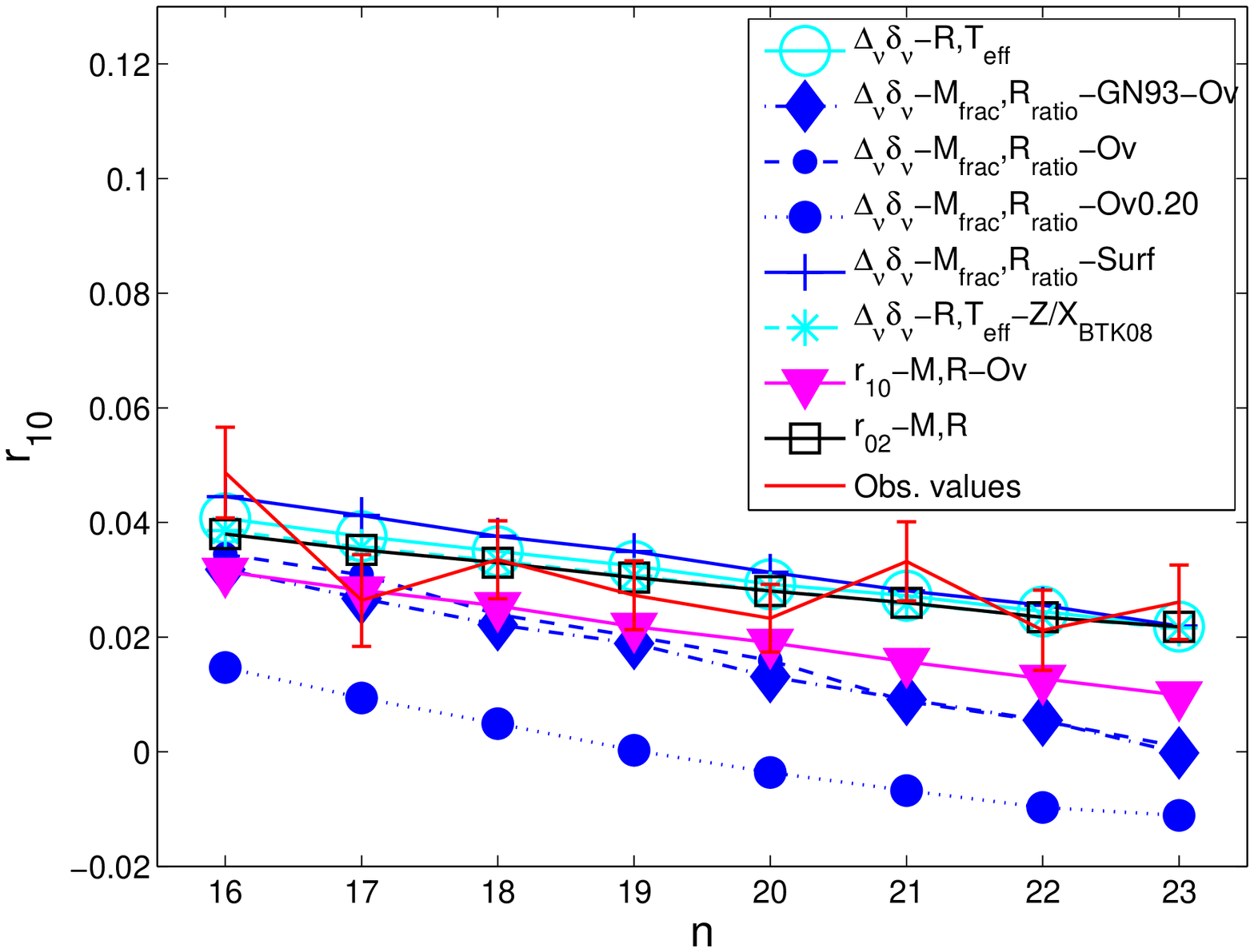}
        \caption{Frequency ratios r$_{02}$ (left panel) and r$_{10}$ (right panel) for a selection of $\alpha$ Cen A 
models from the asteroseismic calibrations. The legend to both plots is given in the right panel.}
                \label{figratiosA}
\end{figure*}

We show  in Sect.~\ref{subconstr} that the resulting models give a wide  range of ages from
$\sim$5.6 to $\sim$8.7~Gyr depending on the set of seismic constraints. Calibrations with
frequency ratios as seismic indicators predict the youngest ages, $\lesssim$6.35~Gyr. Some of the 
$\alpha$~Cen~A models that we obtained also present a convective core, but  only in the cases where 
overshooting was 
included. We investigate the fitting accuracy of the models in more detail with help of r$_{02}$ and r$_{10}$, which 
are 
shown in Fig.~\ref{figratiosA}. We compare four of the models with a convective core 
(A-$\Delta_{\nu}\delta_{\nu}$-M$_{\rm frac}$,R$_{\rm ratio}$-Ov and A-$\Delta_{\nu}\delta_{\nu}$-M$_{\rm frac}$,R$_{\rm 
ratio}$-Ov0.20, 
A-$\Delta_{\nu}\delta_{\nu}$-M$_{\rm frac}$,R$_{\rm ratio}$-GN93-Ov, A-r$_{10}$-M,R-Ov) and a selection of models 
representative 
of the different resulting ages.

Looking at r$_{02}$ in the left panel of Fig.~\ref{figratiosA}, none of the selected models can reproduce all of the 
observed values. The model A-$\Delta_{\nu}\delta_{\nu}$-M$_{\rm frac}$,R$_{\rm ratio}$-Ov, with 
a 
convective 
core, is discarded by this indicator since it reproduces none of the observed values. The peak of r$_{02}$ at $n=16$ 
cannot be reproduced by any of the depicted models. Similarly, the observed dip of r$_{02}$ at $n=23$ is a feature that 
it hardly reproduced. Three of these models,  including one with a convective core 
(A-$\Delta_{\nu}\delta_{\nu}$-M$_{\rm frac}$,R$_{\rm ratio}$-GN93-Ov),  fit it, but then fail to fit the values at other 
$n$ 
orders. 
Since none of the models shows such a dip and    all of them are within $2\sigma$ of this value, the importance 
given to this signature has to be tempered. If we exclude this dip in the analysis, reproducing the 
overall behaviour 
of r$_{02}$ tends to favour models without convective core, except the A-r$_{10}$-M,R-Ov, which has one.

\begin{figure*}[!]
        \centering
                \includegraphics[width=8cm]{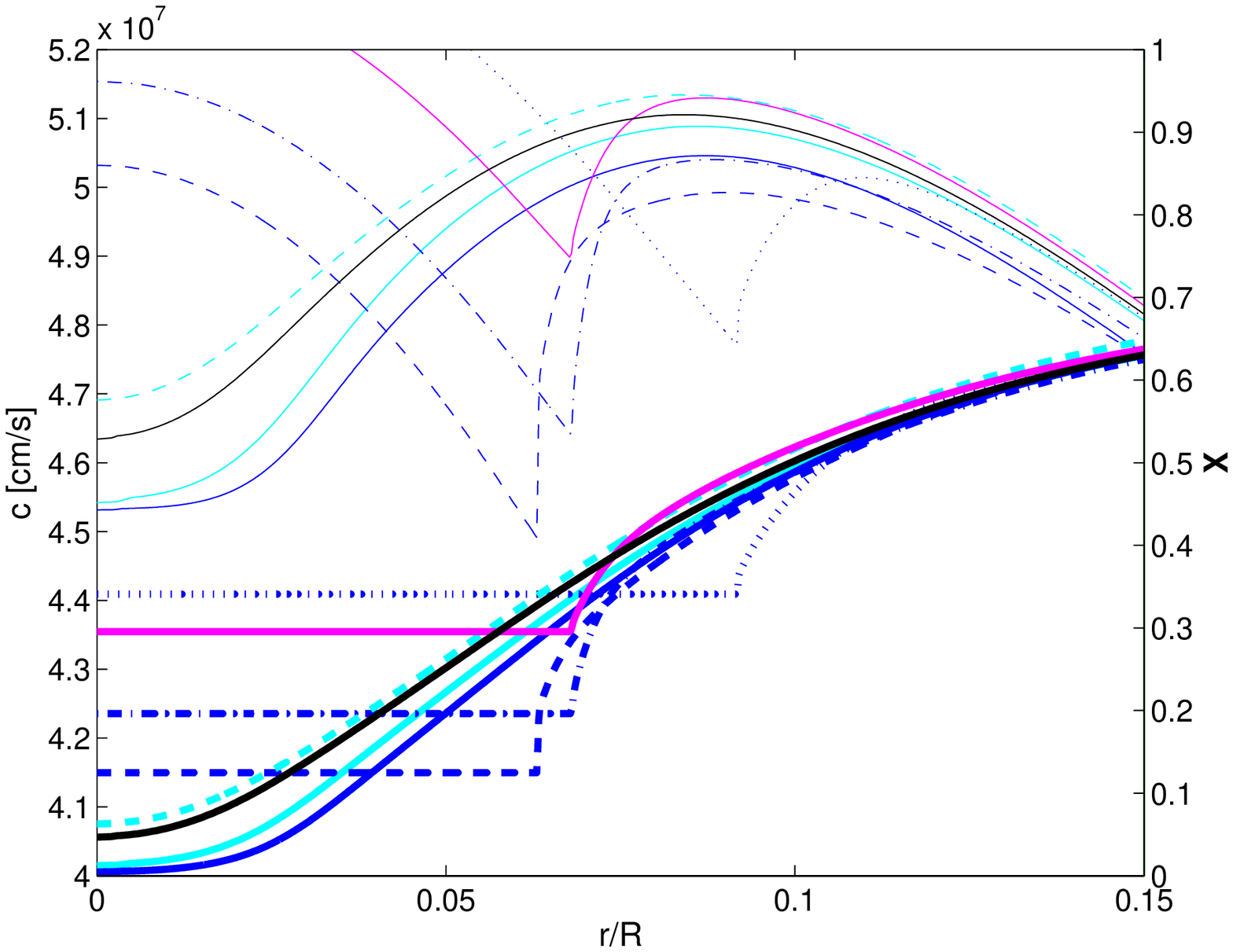}\includegraphics[width=8cm]{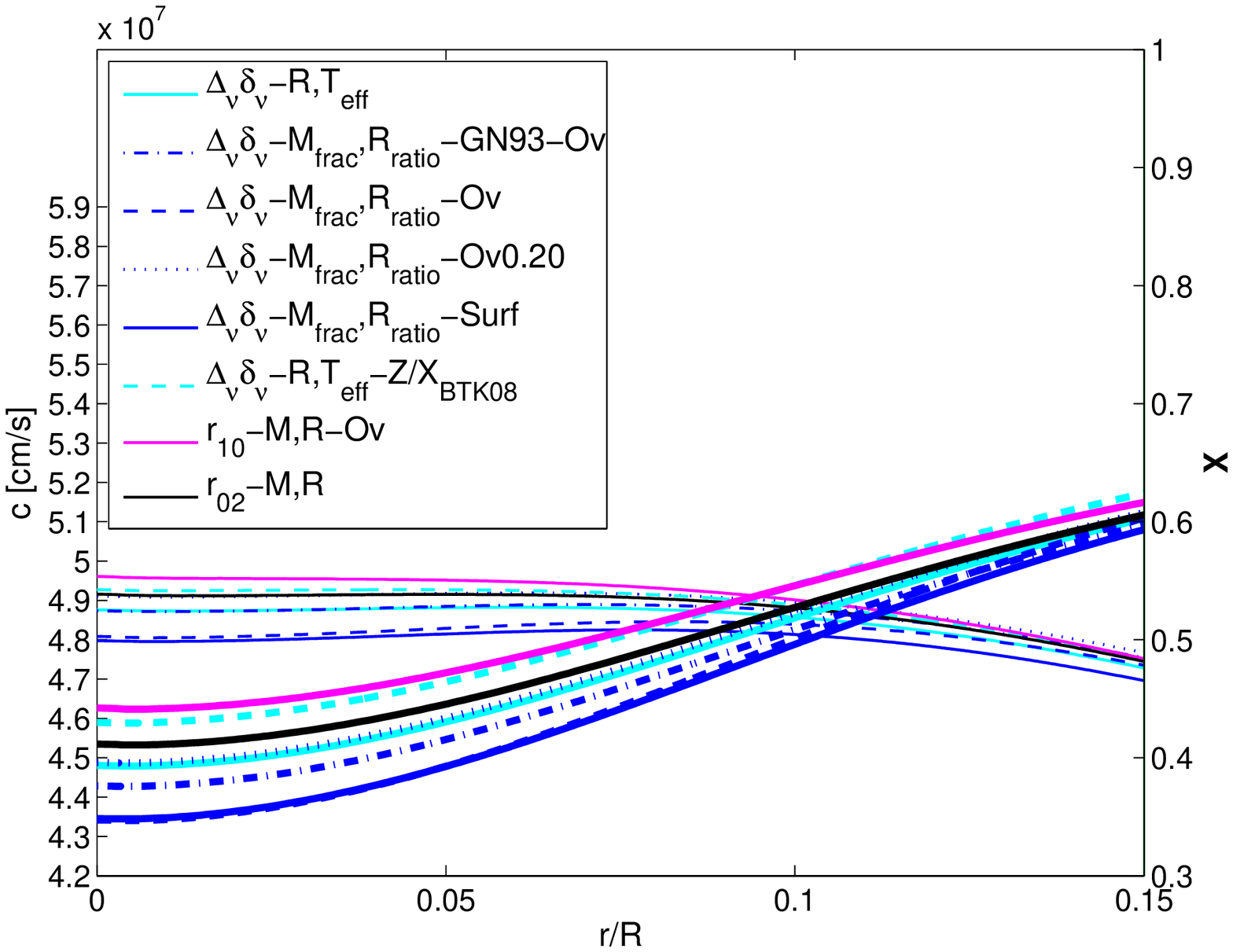}
        \caption{Sound speed (thin lines) and hydrogen abundance (thick lines) for various calibration models of $\alpha$ 
Cen A (left panel) and B (right panel). The legend to both plots is given in the right panel.}
                \label{fig-cx}
\end{figure*}

The internal structure profiles of all these models are shown in the left panel of Fig.~\ref{fig-cx}. The models with a 
convective core have a marked inflection point in their sound speed ($c$) profile close to the centre because of the 
homogeneous chemical composition due to mixing by convection. Models without a convective core show a more or less sharp 
gradient of chemical composition depending on their state of evolution. This leads to a more pronounced contrast 
between 
the central and maximum (at normalised radius r/R$\sim$0.08) values of $c$. The sensitivity of r$_{02}$ to 
stratification in those layers can thus deliver an indication on the age of the star. As seen in the left panel of 
Fig.~\ref{figratiosA}, and referring to Table~\ref{table-results}, models younger than $\sim$6.35~Gyr are 
indeed those that best reproduce this indicator, although a $\sim$7 Gyr model 
(A-$\Delta_{\nu}\delta_{\nu}$-R,T$_{\textrm{eff}}$) is compatible at the margin. The three models that were clearly excluded 
(A-$\Delta_{\nu}\delta_{\nu}$-M$_{\rm frac}$,R$_{\rm ratio}$-GN93-Ov, A-$\Delta_{\nu}\delta_{\nu}$-M$_{\rm 
frac}$,R$_{\rm ratio}$-Ov, 
and 
A-$\Delta_{\nu}\delta_{\nu}$-M$_{\rm frac}$,R$_{\rm ratio}$-Surf) are also the oldest of the selection, respectively 
of 7.20, 
8.12, 
and 8.66~Gyr. 

Focusing on the r$_{10}$ indicator (right panel of Fig.~\ref{figratiosA}), the model with the largest overshoot, 
$\alpha_{\textrm{ov}}=0.2$, is clearly disqualified.  The other models with a convective core also seem to be excluded 
due to their inability to reproduce r$_{10}$ for $n=18$ to 23 ($n=21$ to 23 for the A-r$_{10}$-M,R-Ov model). 

At $n=17$, models with a convective core preferentially fit the observed dip, while only models without one can 
reproduce the peak observed at $n=21$.  However, as none of the models present these oscillating patterns in their r$_{10}$ 
values, we must include with caution the importance given to these features in our analysis. 

Considering the quality of the global fit of these indicators, it seems it favours the absence of a convective core in 
$\alpha$~Cen~A. Nevertheless this conclusion is not definitive since the models without a convective core do not 
reproduce all of the features observed in r$_{02}$ and r$_{01}$. 

Meanwhile, the three oldest models (7.20, 8.12, and 8.66~Gyr) should be discarded because of  the r$_{02}$ indicator. 
The models of 7.20 and 8.66~Gyr are also disqualified by the r$_{10}$ indicator, while it is    less clear for the 
8.12~Gyr model.  The r$_{10}$ indicator also seems  to discard the youngest model (5.55~Gyr; A-r$_{10}$-M,R-Ov), while 
ruling it out from the fit of r$_{02}$ is less clear. The most certain range that emerges is between 
$\sim$6.2 and $\sim$7~Gyr following the present analysis based on the frequency ratios. 

This range is in good agreement with the asteroseismic calibrations of \citet{eggenberger04} and \citet{miglio05}, 
which derived models of between 5.5 and 7 Gyr in age. Our results also corroborate the recent determination by 
\citet{morel18}, 
which derives an age of $\sim$6 Gyr, based on the surface abundances of given chemical species.

\subsubsection{$\alpha$ Cen B}

\begin{figure}[!]
        \centering
                \includegraphics[scale=0.45]{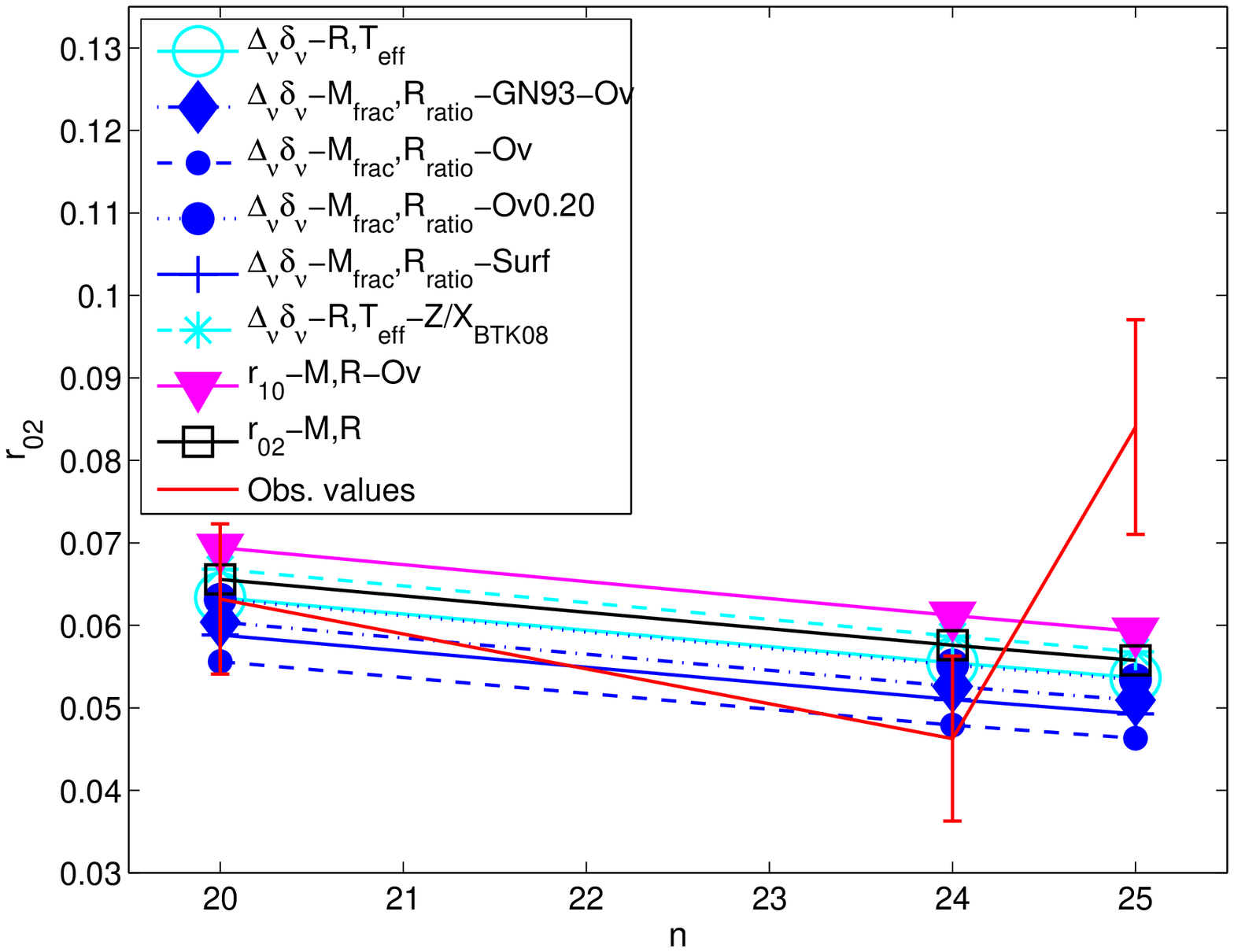}
        \caption{Frequency ratios r$_{02}$ for a selection of $\alpha$ Cen B models from the asteroseismic 
calibrations.}
                \label{figratiosB}
\end{figure}

The analysis of $\alpha$~Cen~B with help of the frequency ratios is limited. Since we have a smaller number of 
frequencies observed for the secondary star, and with less precision, we have for instance only three measurements of r$_{02}$ with larger 
error bars, as shown in Fig.~\ref{figratiosB}.

As a consequence,  r$_{02}$ is difficult to use for retrieving information on the structure of the B component. For 
instance, the value of r$_{02}$ at $n=20$ is reproduced by all of the depicted $\alpha$~Cen~B models in 
Fig.~\ref{figratiosB}. None of the models is able to reproduce the observed value for the 
order $n=25$ of r$_{02}$, although this value looks like an outlier. We note an exception for $n=24$. A look at 
Fig.~\ref{fig-cx} (right panel) of the central chemical composition ($X$) and sound speed ($c$) profiles of a selection 
of 
$\alpha$~Cen~B models shows that models not reproducing r$_{02}$ at this order are those with the highest values of $X$ 
and $c$ at the centre. It suggests that this is a  possible way to estimate a threshold value of $X$ or $c$ at the centre. However, given 
the error on r$_{02}$ and the non-reproduction of it at $n=25$, it is impossible to firmly assert it.

\section{Seismic inversions}
\label{section-inv-seismo}
\begin{figure}[t]
        \centering
                \includegraphics[width=8cm]{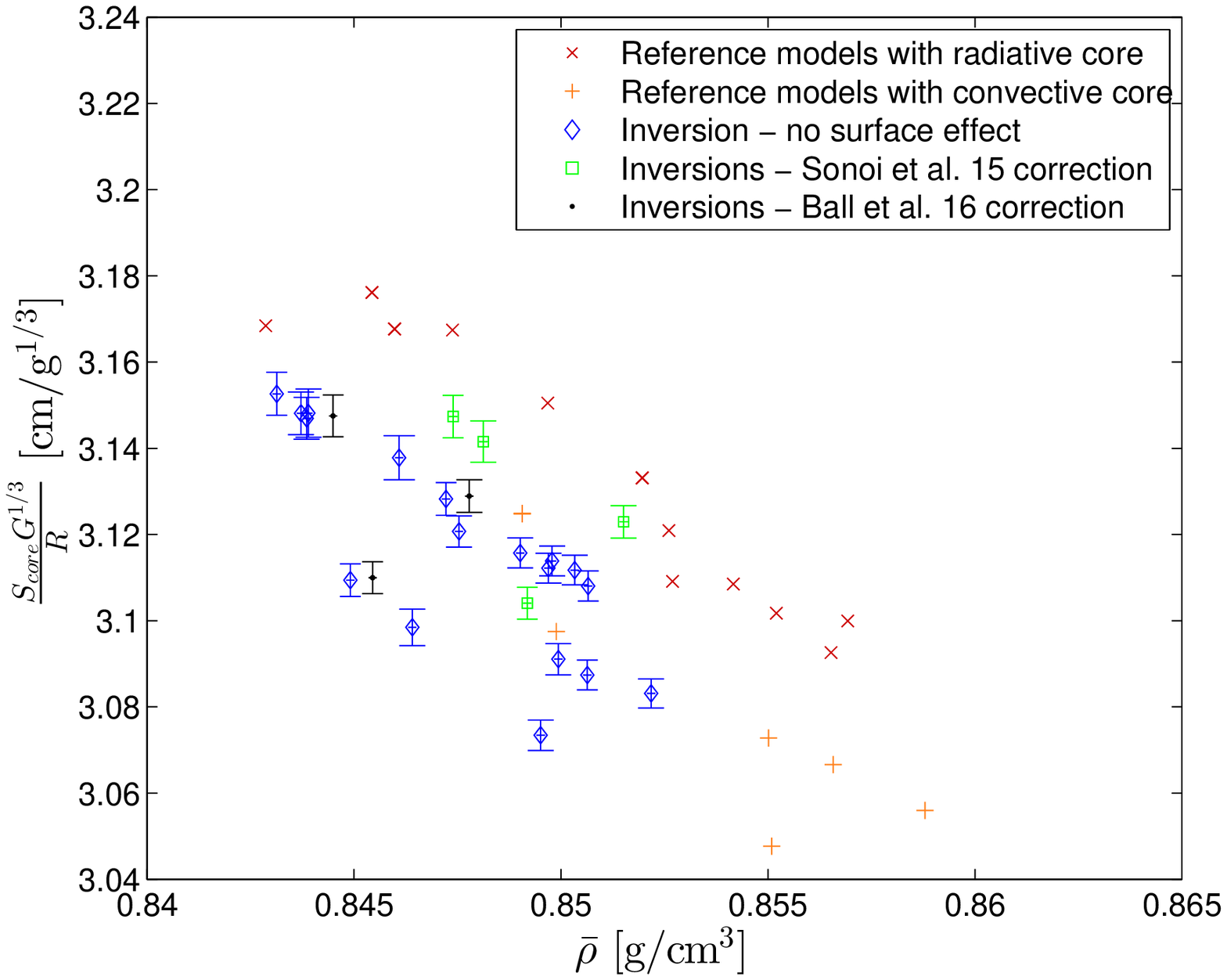}
        \caption{Inversion results for $\alpha$~Cen~A for both the mean density and the $S_{\mathrm{Core}}$ 
indicator. 
Reference values from forward seismic modelling are given in red and orange, corresponding respectively to models 
without or with a convective core. Blue, green and black symbols represent the inversion results obtained 
with different treatments of surface effects (see legend).}
                \label{figInvA}
\end{figure}

\begin{figure}[t]
        \centering
                \includegraphics[width=8cm]{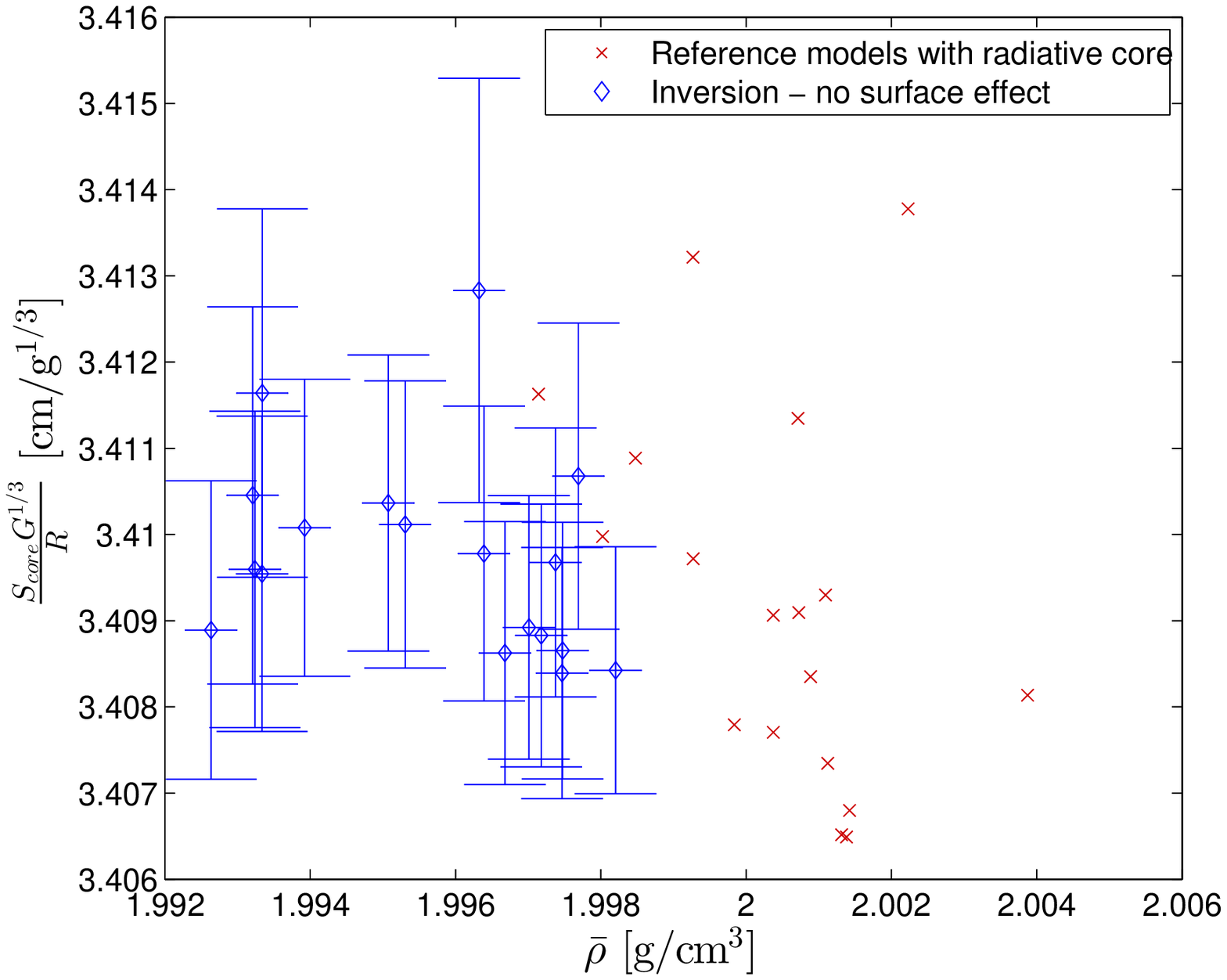}
        \caption{Inversions results for $\alpha$~Cen~B for both the mean density and the $S_{\mathrm{Core}}$ indicator. 
Reference values 
from forward seismic modelling are given in red and inversion results in blue.}
                \label{figInvB}
\end{figure}

The inversion results are presented in Fig.~\ref{figInvA} for $\alpha$~Cen~A and in Fig.~\ref{figInvB} for 
$\alpha$~Cen~B, illustrating the reference and inverted values for the mean density and $S_{\mathrm{Core}}$ indicators 
for various reference models. We used multiple reference models 
fitted using different seismic constraints to take into account the model dependency of the inversion results. As  can 
be seen from Fig.~\ref{figInvA}, this effect is the most significant contributor to the error budget. 

The surface effect corrections of \citet{ball14} were also tested in addition to that of \citet{sonoi15},  used to 
calibrate some reference models in the previous section. The \citet{ball14} correction was implemented directly in the 
SOLA cost function of the inversion, while the \citet{sonoi15} correction was 
implemented using their empirical law in $\log g$ and $T_{\textrm{eff}}$. The surface effects, in particular those 
treated following \citet{sonoi15}, have a significant but slightly less important impact than model dispersion. They 
actually do not increase the overall spread of the inversion results. Hence, for $\alpha$~Cen~B, we only present 
inversion results without surface corrections as the total spread, which is the proper measure of the uncertainties, will be covered by the model-dependency effect.

From the inversion results, it appears that the $S_{\mathrm{Core}}$ indicator is unable to firmly distinguish 
between the absence or presence of a convective core; there are indeed models both without or with a convective core in 
agreement with the inverted values. However, the 
inversion clearly rejects most of the models with ages older than 8~Gyr (the four reference models with the highest values of $S_{\textrm{Core}}$ in Fig.~\ref{figInvA}) previously found by the forward  modelling approach. Similarly, the youngest model at $\sim 5.55$~Gyr is also clearly discarded. It essentially favours models with ages between $\sim$5.9~-~7.3~Gyr, whilst  an older model at 8.66~Gyr also appears  compatible with 
$S_{\mathrm{Core}}$ inverted values. The inversion also provides a $1\%$ interval for the mean density values and it appears that half of 
the models are tilted outside of that interval.

It is actually possible to build models with or without a convective core agreeing with each other at the 1$\sigma$ 
level for all individual frequencies. Such models will have extremely similar mean densities and frequency ratios, and 
thus cannot be distinguished by any seismic analysis technique. In these specific cases the inversion, which uses 
recombinations of the individual frequencies, cannot favour one of the two models.

Similar conclusions can be drawn for $\alpha$~Cen~B since in this case the $S_{\mathrm{Core}}$ inversion is not 
constraining. However, the 
mean density inversion  provides similar constraints and shows that models are in slight disagreement with the 
inverted values. 
Overall, this implies that the inverted mean density values can be used to further constrain the models of both stars 
and especially their masses, but that the final decision on whether $\alpha$~Cen~A harbours a convective core will most 
likely require more precise seismic data. 

\section{Conclusion}

We reconsidered the seismic study of the $\alpha~$Cen~AB binary, based on the oscillation frequency set of \citet{demeulenaer10} and the latest analyses of its orbital motion (P16, K16). In addition to a more traditional forward approach based on the large and small frequency separations, and frequency ratios, we also conducted asteroseismic inversions trying to answer the issue of the presence of a convective core in the primary component, $\alpha$~Cen~A.

We first tested how different choices in the input physics of the models were affecting the results of the asteroseismic 
modelling. This is also a way to identify to what extent the current precision of the frequency dataset for 
$\alpha$~Cen~AB can probe the stellar physics of models. Concerning the choice of the opacity data, we barely notice 
differences between the use of OPAL or OPLIB tables. The most influential element is the chemical mixture. We performed 
most of our seismic modelling using the determination of solar abundances by AGSS09; these models present lower metal 
abundances in comparison to previous determinations, such as GN93. For a given metallicity, a solar-like model will see 
its structure mainly altered by differences in the opacity because of reduced contributions from the metals. As a 
result, switching  from GN93 models to AGSS09 models, we have observed a slight decrease in the masses deduced
from asteroseismology of $\sim$0.02~M$_{\odot}$ for both components.

We also compared these results with those of the orbital solutions by K16 and P16. The asteroseismic masses we obtained with GN93 fall between the K16 and P16 values, a bit closer to K16 than P16. The values obtained with AGSS09 are lower than those derived from the orbital solution by K16, but remain in agreement. They  appear more in disagreement with the P16 orbital solution, however, in particular for the mass of the secondary component. We found one possible reason for this disagreement could be an error in  accuracy on the parallax in P16. However, it also appeared worth questioning the adequacy of adopting solar-distributed abundances in stellar models of $\alpha$~Cen~AB. With the determination of certain photospheric abundances made for this binary \citep[e.g.][]{morel18}, we could consider in the future building composition tables (and the associated opacity and equation of state tables) customised to the analysis of these two stars.

In parallel, we also focused on the age of the system, which we derived to be between 5.9 and 7.3~Gyr, 
regardless of the chemical composition adopted. This estimate is in good agreement with previous asteroseismic values of between 5.5 and 7~Gyr 
\citep[e.g.][]{eggenberger04,miglio05}. More recently, \citet{morel18} estimated a similar age 
for the system, of $\sim$6 Gyr, based on an analysis of the photospheric abundances.

We finally looked at the presence of a convective core in $\alpha$~Cen~A with help of the frequency ratios r$_{02}$ and r$_{10}$. Models with a convective core tend to be excluded by r$_{10}$. This trend for excluding the presence of a convective core for models with AGSS09 differs from the conclusion of a recent study by \citet{nsamba19}. On the contrary, these authors found seismic solutions with a convective core in most cases, whatever  chemical mixture was selected. 

We went a step further to resolve this challenging issue by attempting asteroseismic inversions of the mean density and 
an entropy indicator \citep[see details on the method in][]{buldgenmeand,buldgenS}. The inversions converged for each of 
the inverted quantities, but the precision was insufficient to conclude firmly. While the inversions of the entropy 
indicator favour models without a convective core, some of the models with a convective core remain compatible with the 
inverted values. These first inversions are nevertheless encouraging; with a gain in accuracy on oscillation 
frequencies, inversions could leave no doubt. This gain in accuracy would also be beneficial for $\alpha$~Cen~B, on 
which we have attempted the same inversions; however, the results are presently inconclusive. 

This new study shows how privileged and essential  a system like $\alpha~$Cen~AB is for studying the physics of 
solar-like stars. It demonstrates again the role of asteroseismology to determine fundamental stellar parameters and how 
it can highlight potential flaws in physical or observational data. The current precision on the oscillation frequency 
dataset for $\alpha~$Cen is now the main limiting factor for finer asteroseismic analysis. Achieving at least the same 
precision as that obtained on the 16~Cyg binary by the Kepler satellite appears a reasonable goal for such a bright 
system. However, its magnitude is paradoxically  what hampers it for most   observational facilities. New 
progress could be soon reached with the SONG project, and its extension to the south, a network of spectrographs 
specially designed for observing solar-like oscillations \citep[e.g.][]{song}. Alternatively, the recent development of 
missions based on nanosatellites, a type with very moderate costs and reduced lifespan, offers an opportunity to solve 
this problem. The improvement of $\alpha$~Cen seismic data could indeed  fit a dedicated nanosatellite 
project very
well.

\begin{acknowledgements}
S.J.A.J.S. and P.E have received funding from the European Research Council (ERC) under the European Unions Horizon 2020 
research and innovation programme (grant agreement No 833925, project STAREX). V.V.G is a F.R.S-FNRS Research Associate. 
G.B. acknowledges fundings from the SNF AMBIZIONE grant No 185805 (Seismic inversions and modelling of transport 
processes in stars). 
\end{acknowledgements}

%
\bibliographystyle{aa} 
\bibliography{biblio-alphacen} 

\begin{thebibliography}{92}
\expandafter\ifx\csname natexlab\endcsname\relax\def\natexlab#1{#1}\fi

\bibitem[{{Adelberger} {et~al.}(2011){Adelberger}, {Garc{\'{\i}}a},
  {Robertson}, {Snover}, {Balantekin}, {Heeger}, {Ramsey-Musolf}, {Bemmerer},
  {Junghans}, {Bertulani}, {Chen}, {Costantini}, {Prati}, {Couder},
  {Uberseder}, {Wiescher}, {Cyburt}, {Davids}, {Freedman}, {Gai}, {Gazit},
  {Gialanella}, {Imbriani}, {Greife}, {Hass}, {Haxton}, {Itahashi}, {Kubodera},
  {Langanke}, {Leitner}, {Leitner}, {Vetter}, {Winslow}, {Marcucci},
  {Motobayashi}, {Mukhamedzhanov}, {Tribble}, {Nollett}, {Nunes}, {Park},
  {Parker}, {Schiavilla}, {Simpson}, {Spitaleri}, {Strieder}, {Trautvetter},
  {Suemmerer}, \& {Typel}}]{adelberger11}
{Adelberger}, E.~G., {Garc{\'{\i}}a}, A., {Robertson}, R.~G.~H., {et~al.} 2011,
  Reviews of Modern Physics, 83, 195

\bibitem[{{Angulo} {et~al.}(1999){Angulo}, {Arnould}, {Rayet}, {Descouvemont},
  {Baye}, {Leclercq-Willain}, {Coc}, {Barhoumi}, {Aguer}, {Rolfs}, {Kunz},
  {Hammer}, {Mayer}, {Paradellis}, {Kossionides}, {Chronidou}, {Spyrou},
  {degl'Innocenti}, {Fiorentini}, {Ricci}, {Zavatarelli}, {Providencia},
  {Wolters}, {Soares}, {Grama}, {Rahighi}, {Shotter}, \& {Lamehi
  Rachti}}]{angulo99}
{Angulo}, C., {Arnould}, M., {Rayet}, M., {et~al.} 1999, Nuclear Physics A,
  656, 3

\bibitem[{{Asplund} {et~al.}(2005){Asplund}, {Grevesse}, \&
  {Sauval}}]{asplund05}
{Asplund}, M., {Grevesse}, N., \& {Sauval}, A.~J. 2005, in Astronomical Society
  of the Pacific Conference Series, Vol. 336, Cosmic Abundances as Records of
  Stellar Evolution and Nucleosynthesis, ed. T.~G. {Barnes}, III \& F.~N.
  {Bash}, 25

\bibitem[{{Asplund} {et~al.}(2009){Asplund}, {Grevesse}, {Sauval}, \&
  {Scott}}]{asplund09}
{Asplund}, M., {Grevesse}, N., {Sauval}, A.~J., \& {Scott}, P. 2009, \araa, 47,
  481

\bibitem[{{Baglin} {et~al.}(2006){Baglin}, {Auvergne}, {Barge}, {Deleuil},
  {Catala}, {Michel}, {Weiss}, \& {COROT Team}}]{baglin06}
{Baglin}, A., {Auvergne}, M., {Barge}, P., {et~al.} 2006, in ESA Special
  Publication, Vol. 1306, The CoRoT Mission Pre-Launch Status - Stellar
  Seismology and Planet Finding, ed. M.~{Fridlund}, A.~{Baglin}, J.~{Lochard},
  \& L.~{Conroy}, 33

\bibitem[{Bahcall {et~al.}(2005)Bahcall, Basu, Pinsonneault, \&
  Serenelli}]{bahcall05}
Bahcall, J.~N., Basu, S., Pinsonneault, M., \& Serenelli, A.~M. 2005, The
  Astrophysical Journal, 618, 1049

\bibitem[{{Ball} \& {Gizon}(2014)}]{ball14}
{Ball}, W.~H. \& {Gizon}, L. 2014, \aap, 568, A123

\bibitem[{{Basu} \& {Antia}(2008)}]{basu08}
{Basu}, S. \& {Antia}, H.~M. 2008, \physrep, 457, 217

\bibitem[{{Bazot} {et~al.}(2007){Bazot}, {Bouchy}, {Kjeldsen}, {Charpinet},
  {Laymand}, \& {Vauclair}}]{bazot07}
{Bazot}, M., {Bouchy}, F., {Kjeldsen}, H., {et~al.} 2007, \aap, 470, 295

\bibitem[{{Bazot} {et~al.}(2012){Bazot}, {Bourguignon}, \&
  {Christensen-Dalsgaard}}]{bazot12}
{Bazot}, M., {Bourguignon}, S., \& {Christensen-Dalsgaard}, J. 2012, \mnras,
  427, 1847

\bibitem[{{Bazot} {et~al.}(2016){Bazot}, {Christensen-Dalsgaard}, {Gizon}, \&
  {Benomar}}]{bazot16}
{Bazot}, M., {Christensen-Dalsgaard}, J., {Gizon}, L., \& {Benomar}, O. 2016,
  \mnras, 460, 1254

\bibitem[{{Bedding} \& {Kjeldsen}(2008)}]{bedding08}
{Bedding}, T.~R. \& {Kjeldsen}, H. 2008, in Astronomical Society of the Pacific
  Conference Series, Vol. 384, 14th Cambridge Workshop on Cool Stars, Stellar
  Systems, and the Sun, ed. G.~{van Belle}, 21

\bibitem[{{Bedding} {et~al.}(2004){Bedding}, {Kjeldsen}, {Butler}, {McCarthy},
  {Marcy}, {O'Toole}, {Tinney}, \& {Wright}}]{bedding04}
{Bedding}, T.~R., {Kjeldsen}, H., {Butler}, R.~P., {et~al.} 2004, \apj, 614,
  380

\bibitem[{{Bevington} \& {Robinson}(2003)}]{bevington03}
{Bevington}, P.~R. \& {Robinson}, D.~K. 2003, {Data reduction and error
  analysis for the physical sciences}

\bibitem[{{Bigot} {et~al.}(2008){Bigot}, {Th{\'e}venin}, \&
  {Kervella}}]{bigot08}
{Bigot}, L., {Th{\'e}venin}, F., \& {Kervella}, P. 2008, \memsai, 79, 670

\bibitem[{{B{\"o}hm-Vitense}(1958)}]{bohm58}
{B{\"o}hm-Vitense}, E. 1958, \zap, 46, 108

\bibitem[{{Borucki} {et~al.}(2010){Borucki}, {Koch}, {Basri}, {Batalha},
  {Brown}, {Caldwell}, {Caldwell}, {Christensen-Dalsgaard}, {Cochran},
  {DeVore}, {Dunham}, {Dupree}, {Gautier}, {Geary}, {Gilliland}, {Gould},
  {Howell}, {Jenkins}, {Kondo}, {Latham}, {Marcy}, {Meibom}, {Kjeldsen},
  {Lissauer}, {Monet}, {Morrison}, {Sasselov}, {Tarter}, {Boss}, {Brownlee},
  {Owen}, {Buzasi}, {Charbonneau}, {Doyle}, {Fortney}, {Ford}, {Holman},
  {Seager}, {Steffen}, {Welsh}, {Rowe}, {Anderson}, {Buchhave}, {Ciardi},
  {Walkowicz}, {Sherry}, {Horch}, {Isaacson}, {Everett}, {Fischer}, {Torres},
  {Johnson}, {Endl}, {MacQueen}, {Bryson}, {Dotson}, {Haas}, {Kolodziejczak},
  {Van Cleve}, {Chandrasekaran}, {Twicken}, {Quintana}, {Clarke}, {Allen},
  {Li}, {Wu}, {Tenenbaum}, {Verner}, {Bruhweiler}, {Barnes}, \&
  {Prsa}}]{borucki10}
{Borucki}, W.~J., {Koch}, D., {Basri}, G., {et~al.} 2010, Science, 327, 977

\bibitem[{{Bouchy} \& {Carrier}(2001)}]{bouchy01}
{Bouchy}, F. \& {Carrier}, F. 2001, \aap, 374, L5

\bibitem[{{Bouchy} \& {Carrier}(2002)}]{bouchy02}
{Bouchy}, F. \& {Carrier}, F. 2002, \aap, 390, 205

\bibitem[{{Buldgen} {et~al.}(2016{\natexlab{a}}){Buldgen}, {Reese}, \&
  {Dupret}}]{buldgen16a}
{Buldgen}, G., {Reese}, D.~R., \& {Dupret}, M.~A. 2016{\natexlab{a}}, \aap,
  585, A109

\bibitem[{{Buldgen} {et~al.}(2018){Buldgen}, {Reese}, \& {Dupret}}]{buldgenS}
{Buldgen}, G., {Reese}, D.~R., \& {Dupret}, M.~A. 2018, \aap, 609, A95

\bibitem[{{Buldgen} {et~al.}(2015){Buldgen}, {Reese}, {Dupret}, \&
  {Samadi}}]{buldgenmeand}
{Buldgen}, G., {Reese}, D.~R., {Dupret}, M.~A., \& {Samadi}, R. 2015, \aap,
  574, A42

\bibitem[{Buldgen {et~al.}(2019)Buldgen, Salmon, \& Noels}]{buldgen19}
Buldgen, G., Salmon, S., \& Noels, A. 2019, Frontiers in Astronomy and Space
  Sciences, 6, 42

\bibitem[{{Buldgen} {et~al.}(2016{\natexlab{b}}){Buldgen}, {Salmon}, {Reese},
  \& {Dupret}}]{buldgen16b}
{Buldgen}, G., {Salmon}, S.~J.~A.~J., {Reese}, D.~R., \& {Dupret}, M.~A.
  2016{\natexlab{b}}, \aap, 596, A73

\bibitem[{{Butler} {et~al.}(2004){Butler}, {Bedding}, {Kjeldsen}, {McCarthy},
  {O'Toole}, {Tinney}, {Marcy}, \& {Wright}}]{butler04}
{Butler}, R.~P., {Bedding}, T.~R., {Kjeldsen}, H., {et~al.} 2004, \apjl, 600,
  L75

\bibitem[{{Caffau} {et~al.}(2011){Caffau}, {Ludwig}, {Steffen}, {Freytag}, \&
  {Bonifacio}}]{caffau11}
{Caffau}, E., {Ludwig}, H.-G., {Steffen}, M., {Freytag}, B., \& {Bonifacio}, P.
  2011, \solphys, 268, 255

\bibitem[{{Carrier} \& {Bourban}(2003)}]{carrier03}
{Carrier}, F. \& {Bourban}, G. 2003, \aap, 406, L23

\bibitem[{{Chaplin} \& {Miglio}(2013)}]{chaplin13}
{Chaplin}, W.~J. \& {Miglio}, A. 2013, \araa, 51, 353

\bibitem[{{Colgan} {et~al.}(2016){Colgan}, {Kilcrease}, {Magee}, {Sherrill},
  {Abdallah}, {Hakel}, {Fontes}, {Guzik}, \& {Mussack}}]{colgan16}
{Colgan}, J., {Kilcrease}, D.~P., {Magee}, N.~H., {et~al.} 2016, \apj, 817, 116

\bibitem[{{Cox} \& {Giuli}(1968)}]{cox68}
{Cox}, J.~P. \& {Giuli}, R.~T. 1968, {Principles of stellar structure }

\bibitem[{{de Meulenaer} {et~al.}(2010){de Meulenaer}, {Carrier}, {Miglio},
  {Bedding}, {Campante}, {Eggenberger}, {Kjeldsen}, \&
  {Montalb{\'a}n}}]{demeulenaer10}
{de Meulenaer}, P., {Carrier}, F., {Miglio}, A., {et~al.} 2010, \aap, 523, A54

\bibitem[{{Demory} {et~al.}(2015){Demory}, {Ehrenreich}, {Queloz}, {Seager},
  {Gilliland}, {Chaplin}, {Proffitt}, {Gillon}, {G{\"u}nther}, {Benneke},
  {Dumusque}, {Lovis}, {Pepe}, {S{\'e}gransan}, {Triaud}, \& {Udry}}]{demory15}
{Demory}, B.-O., {Ehrenreich}, D., {Queloz}, D., {et~al.} 2015, \mnras, 450,
  2043

\bibitem[{{Ducati}(2002)}]{ducati02}
{Ducati}, J.~R. 2002, VizieR Online Data Catalog, 2237

\bibitem[{{Dumusque} {et~al.}(2012){Dumusque}, {Pepe}, {Lovis},
  {S{\'e}gransan}, {Sahlmann}, {Benz}, {Bouchy}, {Mayor}, {Queloz}, {Santos},
  \& {Udry}}]{dumusque12}
{Dumusque}, X., {Pepe}, F., {Lovis}, C., {et~al.} 2012, \nat, 491, 207

\bibitem[{{Dziembowski} {et~al.}(1990){Dziembowski}, {Pamyatnykh}, \&
  {Sienkiewicz}}]{dziembowski90}
{Dziembowski}, W.~A., {Pamyatnykh}, A.~A., \& {Sienkiewicz}, R. 1990, \mnras,
  244, 542

\bibitem[{{Edmonds} {et~al.}(1992){Edmonds}, {Cram}, {Demarque}, {Guenther}, \&
  {Pinsonneault}}]{edmonds92}
{Edmonds}, P., {Cram}, L., {Demarque}, P., {Guenther}, D.~B., \&
  {Pinsonneault}, M.~H. 1992, \apj, 394, 313

\bibitem[{{Edvardsson}(1988)}]{edvardsson88}
{Edvardsson}, B. 1988, \aap, 190, 148

\bibitem[{{Eggenberger} {et~al.}(2004){Eggenberger}, {Charbonnel}, {Talon},
  {Meynet}, {Maeder}, {Carrier}, \& {Bourban}}]{eggenberger04}
{Eggenberger}, P., {Charbonnel}, C., {Talon}, S., {et~al.} 2004, \aap, 417, 235

\bibitem[{{Ferguson} {et~al.}(2005){Ferguson}, {Alexander}, {Allard}, {Barman},
  {Bodnarik}, {Hauschildt}, {Heffner-Wong}, \& {Tamanai}}]{ferguson05}
{Ferguson}, J.~W., {Alexander}, D.~R., {Allard}, F., {et~al.} 2005, \apj, 623,
  585

\bibitem[{{Fletcher} {et~al.}(2006){Fletcher}, {Chaplin}, {Elsworth}, {Schou},
  \& {Buzasi}}]{fletcher06}
{Fletcher}, S.~T., {Chaplin}, W.~J., {Elsworth}, Y., {Schou}, J., \& {Buzasi},
  D. 2006, \mnras, 371, 935

\bibitem[{{French} \& {Powell}(1971)}]{french71}
{French}, V.~A. \& {Powell}, A.~L.~T. 1971, Royal Greenwich Observatory
  Bulletins, 173, 63

\bibitem[{Gough(2003)}]{Gough2003}
Gough, D. 2003, Astrophysics and Space Science, 284, 165

\bibitem[{{Grevesse} \& {Noels}(1993)}]{grevesse93}
{Grevesse}, N. \& {Noels}, A. 1993, in Origin and Evolution of the Elements,
  ed. N.~{Prantzos}, E.~{Vangioni-Flam}, \& M.~{Casse}, 15--25

\bibitem[{{Grevesse} \& {Sauval}(1998)}]{grevesse98}
{Grevesse}, N. \& {Sauval}, A.~J. 1998, \ssr, 85, 161

\bibitem[{{Grundahl} {et~al.}(2009){Grundahl}, {Christensen-Dalsgaard},
  {Arentoft}, {Frandsen}, {Kjeldsen}, {J{\o}rgensen}, \&
  {Kj{\ae}rgaard}}]{song}
{Grundahl}, F., {Christensen-Dalsgaard}, J., {Arentoft}, T., {et~al.} 2009,
  Communications in Asteroseismology, 158, 345

\bibitem[{{Guenther} \& {Demarque}(2000)}]{guenther00}
{Guenther}, D.~B. \& {Demarque}, P. 2000, \apj, 531, 503

\bibitem[{{Hatzes}(2013)}]{hatzes13}
{Hatzes}, A.~P. 2013, \apj, 770, 133

\bibitem[{{Iglesias} \& {Rogers}(1996)}]{iglesias96}
{Iglesias}, C.~A. \& {Rogers}, F.~J. 1996, \apj, 464, 943

\bibitem[{{Irwin}(2012)}]{irwin12}
{Irwin}, A.~W. 2012, {FreeEOS: Equation of State for stellar interiors
  calculations}, Astrophysics Source Code Library

\bibitem[{{Joyce} \& {Chaboyer}(2018)}]{joyce18}
{Joyce}, M. \& {Chaboyer}, B. 2018, \apj, 864, 99

\bibitem[{{Kaltenegger} \& {Haghighipour}(2013)}]{kaltenegger13}
{Kaltenegger}, L. \& {Haghighipour}, N. 2013, \apj, 777, 165

\bibitem[{{Kamper} \& {Wesselink}(1978)}]{kamper78}
{Kamper}, K.~W. \& {Wesselink}, A.~J. 1978, \aj, 83, 1653

\bibitem[{{Kervella} {et~al.}(2017{\natexlab{a}}){Kervella}, {Bigot},
  {Gallenne}, \& {Th{\'e}venin}}]{kervella17}
{Kervella}, P., {Bigot}, L., {Gallenne}, A., \& {Th{\'e}venin}, F.
  2017{\natexlab{a}}, \aap, 597, A137

\bibitem[{{Kervella} {et~al.}(2016){Kervella}, {Mignard}, {M{\'e}rand}, \&
  {Th{\'e}venin}}]{kervella16}
{Kervella}, P., {Mignard}, F., {M{\'e}rand}, A., \& {Th{\'e}venin}, F. 2016,
  \aap, 594, A107

\bibitem[{{Kervella} {et~al.}(2017{\natexlab{b}}){Kervella}, {Th{\'e}venin}, \&
  {Lovis}}]{kervella17b}
{Kervella}, P., {Th{\'e}venin}, F., \& {Lovis}, C. 2017{\natexlab{b}}, \aap,
  598, L7

\bibitem[{{Kervella} {et~al.}(2003){Kervella}, {Th{\'e}venin}, {S{\'e}gransan},
  {Berthomieu}, {Lopez}, {Morel}, \& {Provost}}]{kervella03}
{Kervella}, P., {Th{\'e}venin}, F., {S{\'e}gransan}, D., {et~al.} 2003, \aap,
  404, 1087

\bibitem[{{Kim}(1999)}]{kim99}
{Kim}, Y.-C. 1999, Journal of Korean Astronomical Society, 32, 119

\bibitem[{{Kjeldsen} {et~al.}(2005){Kjeldsen}, {Bedding}, {Butler},
  {Christensen-Dalsgaard}, {Kiss}, {McCarthy}, {Marcy}, {Tinney}, \&
  {Wright}}]{kjeldsen05}
{Kjeldsen}, H., {Bedding}, T.~R., {Butler}, R.~P., {et~al.} 2005, \apj, 635,
  1281

\bibitem[{{Kjeldsen} {et~al.}(2008){Kjeldsen}, {Bedding}, \&
  {Christensen-Dalsgaard}}]{kjeldsen08}
{Kjeldsen}, H., {Bedding}, T.~R., \& {Christensen-Dalsgaard}, J. 2008, \apjl,
  683, L175

\bibitem[{{Kjeldsen} {et~al.}(1999){Kjeldsen}, {Bedding}, {Frandsen}, \&
  {dall}}]{kjeldsen99}
{Kjeldsen}, H., {Bedding}, T.~R., {Frandsen}, S., \& {dall}, T.~H. 1999,
  \mnras, 303, 579

\bibitem[{{Le Pennec} {et~al.}(2015){Le Pennec}, {Turck-Chi{\`e}ze}, {Salmon},
  {Blancard}, {Coss{\'e}}, {Faussurier}, \& {Mondet}}]{lepennec15}
{Le Pennec}, M., {Turck-Chi{\`e}ze}, S., {Salmon}, S., {et~al.} 2015, \apjl,
  813, L42

\bibitem[{{Lebreton} \& {Goupil}(2014)}]{lebreton14}
{Lebreton}, Y. \& {Goupil}, M.~J. 2014, \aap, 569, A21

\bibitem[{{Lund} {et~al.}(2017){Lund}, {Silva Aguirre}, {Davies}, {Chaplin},
  {Christensen-Dalsgaard}, {Houdek}, {White}, {Bedding}, {Ball}, {Huber},
  {Antia}, {Lebreton}, {Latham}, {Handberg}, {Verma}, {Basu}, {Casagrande},
  {Justesen}, {Kjeldsen}, \& {Mosumgaard}}]{lund17}
{Lund}, M.~N., {Silva Aguirre}, V., {Davies}, G.~R., {et~al.} 2017, \apj, 835,
  172

\bibitem[{{Metcalfe} {et~al.}(2012){Metcalfe}, {Chaplin}, {Appourchaux},
  {Garc{\'{\i}}a}, {Basu}, {Brand{\~a}o}, {Creevey}, {Deheuvels}, {Do{\v g}an},
  {Eggenberger}, {Karoff}, {Miglio}, {Stello}, {Y{\i}ld{\i}z}, {{\c C}elik},
  {Antia}, {Benomar}, {Howe}, {R{\'e}gulo}, {Salabert}, {Stahn}, {Bedding},
  {Davies}, {Elsworth}, {Gizon}, {Hekker}, {Mathur}, {Mosser}, {Bryson},
  {Still}, {Christensen-Dalsgaard}, {Gilliland}, {Kawaler}, {Kjeldsen},
  {Ibrahim}, {Klaus}, \& {Li}}]{metcalfe12}
{Metcalfe}, T.~S., {Chaplin}, W.~J., {Appourchaux}, T., {et~al.} 2012, \apjl,
  748, L10

\bibitem[{{Miglio} \& {Montalb{\'a}n}(2005)}]{miglio05}
{Miglio}, A. \& {Montalb{\'a}n}, J. 2005, \aap, 441, 615

\bibitem[{{Mondet} {et~al.}(2015){Mondet}, {Blancard}, {Coss{\'e}}, \&
  {Faussurier}}]{mondet15}
{Mondet}, G., {Blancard}, C., {Coss{\'e}}, P., \& {Faussurier}, G. 2015, \apjs,
  220, 2

\bibitem[{{Morel}(2018)}]{morel18}
{Morel}, T. 2018, \aap, 615, A172

\bibitem[{{Neuforge} {et~al.}(1999){Neuforge}, {Pourbaix}, {Noels}, \&
  {Scuflaire}}]{neuforge99}
{Neuforge}, C., {Pourbaix}, D., {Noels}, A., \& {Scuflaire}, R. 1999,
  Astronomical Society of the Pacific Conference Series, Vol. 185, {Upward
  Revision of the Individual Masses in A Cen: Implications for the Evolutionary
  State of the System}, ed. J.~B. {Hearnshaw} \& C.~D. {Scarfe}, 335

\bibitem[{{Neuforge-Verheecke} \& {Magain}(1997)}]{neuforge97}
{Neuforge-Verheecke}, C. \& {Magain}, P. 1997, \aap, 328, 261

\bibitem[{{Nsamba} {et~al.}(2019){Nsamba}, {Campante}, {Monteiro}, {Cunha}, \&
  {Sousa}}]{nsamba19}
{Nsamba}, B., {Campante}, T.~L., {Monteiro}, M. J.~P.~F.~G., {Cunha}, M.~S., \&
  {Sousa}, S.~G. 2019, Frontiers in Astronomy and Space Sciences, 6, 25

\bibitem[{{Nsamba} {et~al.}(2018){Nsamba}, {Monteiro}, {Campante}, {Cunha}, \&
  {Sousa}}]{nsamba18}
{Nsamba}, B., {Monteiro}, M.~J.~P.~F.~G., {Campante}, T.~L., {Cunha}, M.~S., \&
  {Sousa}, S.~G. 2018, \mnras, 479, L55

\bibitem[{{Porto de Mello} {et~al.}(2008){Porto de Mello}, {Lyra}, \&
  {Keller}}]{portodemello08}
{Porto de Mello}, G.~F., {Lyra}, W., \& {Keller}, G.~R. 2008, \aap, 488, 653

\bibitem[{{Pourbaix}(1998)}]{pourbaix98}
{Pourbaix}, D. 1998, \aaps, 131, 377

\bibitem[{{Pourbaix} \& {Boffin}(2016)}]{pourbaix16}
{Pourbaix}, D. \& {Boffin}, H.~M.~J. 2016, \aap, 586, A90

\bibitem[{{Pourbaix} {et~al.}(2002){Pourbaix}, {Nidever}, {McCarthy}, {Butler},
  {Tinney}, {Marcy}, {Jones}, {Penny}, {Carter}, {Bouchy}, {Pepe}, {Hearnshaw},
  {Skuljan}, {Ramm}, \& {Kent}}]{pourbaix02}
{Pourbaix}, D., {Nidever}, D., {McCarthy}, C., {et~al.} 2002, \aap, 386, 280

\bibitem[{{Quarles} \& {Lissauer}(2016)}]{quarles16}
{Quarles}, B. \& {Lissauer}, J.~J. 2016, \aj, 151, 111

\bibitem[{{Rajpaul} {et~al.}(2016){Rajpaul}, {Aigrain}, \&
  {Roberts}}]{rajpaul16}
{Rajpaul}, V., {Aigrain}, S., \& {Roberts}, S. 2016, \mnras, 456, L6

\bibitem[{{Reese} {et~al.}(2012){Reese}, {Marques}, {Goupil}, {Thompson}, \&
  {Deheuvels}}]{reese12}
{Reese}, D.~R., {Marques}, J.~P., {Goupil}, M.~J., {Thompson}, M.~J., \&
  {Deheuvels}, S. 2012, \aap, 539, A63

\bibitem[{{Roxburgh} \& {Vorontsov}(2003)}]{roxburgh03}
{Roxburgh}, I.~W. \& {Vorontsov}, S.~V. 2003, \aap, 411, 215

\bibitem[{{Schou} \& {Buzasi}(2001)}]{schou01}
{Schou}, J. \& {Buzasi}, D.~L. 2001, in ESA Special Publication, Vol. 464, SOHO
  10/GONG 2000 Workshop: Helio- and Asteroseismology at the Dawn of the
  Millennium, ed. A.~{Wilson} \& P.~L. {Pall{\'e}}, 391--394

\bibitem[{{Scuflaire} {et~al.}(2008{\natexlab{a}}){Scuflaire}, {Montalb{\'a}n},
  {Th{\'e}ado}, {Bourge}, {Miglio}, {Godart}, {Thoul}, \& {Noels}}]{losc}
{Scuflaire}, R., {Montalb{\'a}n}, J., {Th{\'e}ado}, S., {et~al.}
  2008{\natexlab{a}}, \apss, 316, 149

\bibitem[{{Scuflaire} {et~al.}(2008{\natexlab{b}}){Scuflaire}, {Th{\'e}ado},
  {Montalb{\'a}n}, {Miglio}, {Bourge}, {Godart}, {Thoul}, \& {Noels}}]{cles}
{Scuflaire}, R., {Th{\'e}ado}, S., {Montalb{\'a}n}, J., {et~al.}
  2008{\natexlab{b}}, \apss, 316, 83

\bibitem[{{Silva Aguirre} {et~al.}(2015){Silva Aguirre}, {Davies}, {Basu},
  {Christensen-Dalsgaard}, {Creevey}, {Metcalfe}, {Bedding}, {Casagrande},
  {Handberg}, {Lund}, {Nissen}, {Chaplin}, {Huber}, {Serenelli}, {Stello}, {Van
  Eylen}, {Campante}, {Elsworth}, {Gilliland}, {Hekker}, {Karoff}, {Kawaler},
  {Kjeldsen}, \& {Lundkvist}}]{sa15}
{Silva Aguirre}, V., {Davies}, G.~R., {Basu}, S., {et~al.} 2015, \mnras, 452,
  2127

\bibitem[{{Sonoi} {et~al.}(2015){Sonoi}, {Samadi}, {Belkacem}, {Ludwig},
  {Caffau}, \& {Mosser}}]{sonoi15}
{Sonoi}, T., {Samadi}, R., {Belkacem}, K., {et~al.} 2015, \aap, 583, A112

\bibitem[{{Tang} {et~al.}(2008){Tang}, {Bi}, {Gai}, \& {Xu}}]{tang08}
{Tang}, Y.-K., {Bi}, S.-L., {Gai}, N., \& {Xu}, H.-Y. 2008, \cjaa, 8, 421

\bibitem[{{Th{\'e}venin} {et~al.}(2002){Th{\'e}venin}, {Provost}, {Morel},
  {Berthomieu}, {Bouchy}, \& {Carrier}}]{thevenin02}
{Th{\'e}venin}, F., {Provost}, J., {Morel}, P., {et~al.} 2002, \aap, 392, L9

\bibitem[{{Thoul} {et~al.}(2003){Thoul}, {Scuflaire}, {Noels}, {Vatovez},
  {Briquet}, {Dupret}, \& {Montalban}}]{thoul03}
{Thoul}, A., {Scuflaire}, R., {Noels}, A., {et~al.} 2003, \aap, 402, 293

\bibitem[{{Thoul} {et~al.}(1994){Thoul}, {Bahcall}, \& {Loeb}}]{thoul94}
{Thoul}, A.~A., {Bahcall}, J.~N., \& {Loeb}, A. 1994, \apj, 421, 828

\bibitem[{{Wesselink}(1953)}]{wesselink53}
{Wesselink}, A.~J. 1953, \mnras, 113, 505

\bibitem[{{Xu} {et~al.}(2013){Xu}, {Takahashi}, {Goriely}, {Arnould}, {Ohta},
  \& {Utsunomiya}}]{xu13}
{Xu}, Y., {Takahashi}, K., {Goriely}, S., {et~al.} 2013, Nuclear Physics A,
  918, 61

\bibitem[{{Y{\i}ld{\i}z}(2007)}]{yildiz07}
{Y{\i}ld{\i}z}, M. 2007, \mnras, 374, 1264

\bibitem[{{Y{\i}ld{\i}z}(2008)}]{yildiz08}
{Y{\i}ld{\i}z}, M. 2008, \mnras, 388, 1143

\end{thebibliography}
%

%
%

\end{document}